\documentclass[10pt,journal,compsoc]{IEEEtran}
\usepackage{multirow}
\usepackage{amssymb}
\usepackage{bbding}
\usepackage{graphicx}
\usepackage{pifont}
\usepackage{amsmath}
\usepackage{algorithm}  
\usepackage{algorithmicx}
\usepackage{algpseudocode}
\usepackage{subfigure}
\usepackage{bbm}
\usepackage{makecell}
\usepackage{dsfont}
\usepackage{threeparttable}
\usepackage{url}

\ifCLASSOPTIONcompsoc
  \usepackage[nocompress]{cite}
\else
  \usepackage{cite}
\fi
\ifCLASSINFOpdf
\else
\fi
\begin{document}
%
\title{WL-Align: Weisfeiler-Lehman Relabeling for Aligning Users across Networks via Regularized Representation Learning}
%
%
%
%

\author{Li~Liu,~
        Penggang~Chen,~
        Xin~Li,~\IEEEmembership{Member,~IEEE},
        William K.~Cheung,~\IEEEmembership{Member,~IEEE},
        Youmin~Zhang,~
        Qun~Liu,~
        and ~Guoyin~Wang,~\IEEEmembership{Senior Member,~IEEE}
\IEEEcompsocitemizethanks{\IEEEcompsocthanksitem Li Liu, Youmin Zhang, Penggang Chen, Qun Liu and Guoyin Wang are with the Chongqing Key Laboratory of
Computational Intelligence, Chongqing University of Posts and Telecommunications, Chongqing 400065, China.\protect\\
Corresponding authors: Xin Li, William K. Cheung. E-mail: xinli@bit.edu.cn, william@comp.hkbu.edu.hk
\IEEEcompsocthanksitem Xin Li is with Beijing Institute of Technology, Beijing, China, 100081.
\IEEEcompsocthanksitem Li Liu and William K. Cheung are with Hong Kong Baptist University, Hong Kong, China.}
\thanks{Manuscript received}}

%
%

\markboth{Journal of \LaTeX\ Class Files,~Vol.~14, No.~8, August~2015}%
{Shell \MakeLowercase{\textit{et al.}}: Bare Demo of IEEEtran.cls for Computer Society Journals}
%



\IEEEtitleabstractindextext{%
\begin{abstract}
Aligning users across networks using graph representation learning has been found effective where the alignment is accomplished in a low-dimensional embedding space. Yet, achieving highly precise alignment is still challenging, especially when nodes with long-range connectivity to the labeled anchors are encountered. To alleviate this limitation, we purposefully designed WL-Align which adopts a regularized representation learning framework to learn distinctive node representations. It extends the Weisfeiler-Lehman Isormorphism Test and learns the alignment in alternating phases of ``across-network Weisfeiler-Lehman relabeling'' and ``proximity-preserving representation learning''. The across-network Weisfeiler-Lehman relabeling is achieved through iterating the anchor-based label propagation and a similarity-based hashing to exploit the known anchors' connectivity to different nodes in an efficient and robust manner. The representation learning module preserves the second-order proximity within individual networks and is regularized by the across-network Weisfeiler-Lehman hash labels. Extensive experiments on real-world and synthetic datasets have demonstrated that our proposed WL-Align outperforms the state-of-the-art methods, achieving significant performance improvements in the ``exact matching'' scenario. Data and code of WL-Align are available at \url{https://github.com/ChenPengGang/WLAlignCode}.

\end{abstract}

\begin{IEEEkeywords}
Network alignment, Weisfeiler-Lehman Test, Representation Learning, Social Networks
\end{IEEEkeywords}}

\maketitle

\IEEEdisplaynontitleabstractindextext

%
\IEEEpeerreviewmaketitle

\IEEEraisesectionheading{\section{Introduction}\label{sec:introduction}}

%
%
%
%
\IEEEPARstart {A}{ligning} users across networks aims to identify the identities of a natural person in multiple networks, which can benefit data mining tasks including user behavior prediction~\cite{DBLP:conf/aaai/JiangCYXY16}, friend recommendation~\cite{DBLP:journals/sigkdd/ShuWTZL16}, and identity verification~\cite{DBLP:conf/www/GogaLPFST13}.
Graph representation learning (GRL) based algorithms~\cite{DBLP:conf/ijcai/LiuCLL16,DBLP:conf/ijcai/ManSLJC16,heimann2018regal,NeXtAlign,DBLP:conf/kdd/Gao0L21,DBLP:journals/tkde/LiuLCL20,DBLP:conf/cikm/XiaG021,DBLP:conf/www/ChuFYZHB19} have demonstrated superior performance for the user alignment task. 
These algorithms derive the user representations via preserving the structural proximity as well as the user content similarity as revealed in the social networks. 
While the user generated contents and profiles are helpful supplementary information for the alignment, extracting them sometimes can be non-trivial due to factors like their heterogeneous modalities and the privacy protection concern.
Therefore, for the most part, structure-based models still play a crucial role in social network alignment.

In practice, only a portion of known common users (also called anchors) can be identified as the ground truth across social networks, and most of the potential anchors are distributed in the network as their ``higher-order'' neighbours.
To guarantee that each node can acquire identifiable information from the topological long-range anchors, GRL algorithms typically need to aggregate ``higher-order'' neighbours to learn discriminative representations of nodes. Nevertheless, while the aggregation can preserve structural proximity, it has also been pointed out that it can lead to an \textit{over-smoothing} representation space~\cite{DBLP:conf/aaai/LiHW18,DBLP:conf/iclr/ZhaoA20}. 
Differentiating nodes in the over-smoothed space is difficult, which in turn will impair the performance of the GRL-based network alignment methods~\cite{DBLP:conf/ijcai/LiuCLL16,DANA,NeXtAlign,DBLP:conf/kdd/Gao0L21,DBLP:journals/tkde/LiuLCL20,DBLP:conf/www/ChuFYZHB19}, especially if highly precise ``exact matching'' is needed.

To this end, we propose to better organize the representation space for user alignment via a novel way that deliberately exploits the information of the anchors to learn more distinguishable representations.
In particular, we are inspired by the Weisfeiler-Lehman (WL) graph isomorphism test~\cite{wl_test,DBLP:conf/iclr/XuHLJ19} which can effectively test the isomorphism (topological equivalence) of two graphs via label propagation and compression.
We extend the idea to characterize nodes in multiple networks by exploring their topological relationships with known anchors. To do this, we introduce an anchor-based label propagation and hashing strategy with the objective to explore the isomorphic sub-graphs across networks to guide the GRL for the alignment.

The key idea is that we first perform label propagation according to the WL Test based on the known anchors that contain identifiable information across the networks. We further use an injective hashing operation to compute for each non-anchor node a ``compressed'' label based on the set of compressed labels of its neighbours. The anchor information will then be propagated to different nodes according to their connectivity patterns.

Our conjecture is that nodes in different networks assigned with the same ``compressed'' label are highly likely to be the potential anchor pairs across networks. Since the compressed label is obtained via the label propagation and the hashing procedures that originated from the known anchors, nodes with the same label indicate that they share similar topological relationships with respect to the known anchors, and thus are more likely to form an anchor pair. 
This can then provide clues on determining if nodes should be closely organized and which should fall apart from each other in the representation space. We then propose WL-Align which adopts a regularized representation learning framework with the objective to preserve both the compressed labels and the structure proximity simultaneously to produce distinguishable node representations for network alignment. 
In contrast to the aggregation schema used in GRL, which leads to embeddings converging to similar values due to the high probability of accessing the same ``higher-order'' neighbours, the one-hot compressed labels obtained via an injective hash operation on the multiset are distinguishable. Preserving the proximity of the compressed label helps our model to reorganize the representation space, mitigating the "over-smoothing" issue.

\begin{figure}
	\centering
	\includegraphics[width=0.45\textwidth]{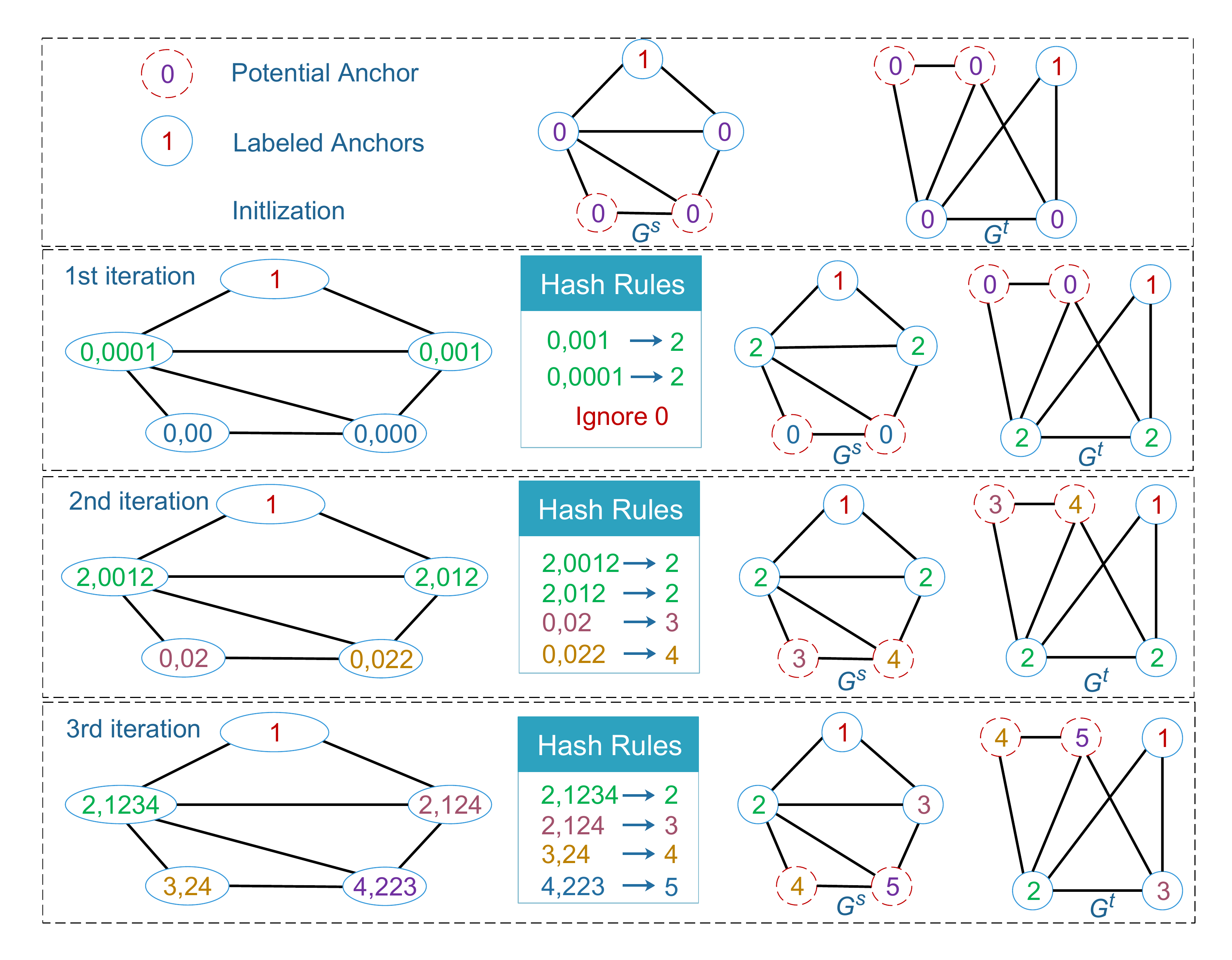}
	\caption{An example of applying WL-Align for user alignment across networks. 
	Note that the anchor labels are kept unique and consistent during iterations. And the "0" label is ignored in the hash rules as it does not provide a discriminative signal to supervise the learning of unified representations
	}
	\label{ToyExamOne}
\end{figure}

Fig. \ref{ToyExamOne} shows an example to illustrate the key ideas of why the WL Test can benefit the alignment across the networks $G^s$ and $G^t$. Initially, the known anchor is assigned with label 1 and the other (potential anchor) nodes are assigned with label 0.\footnote{For cases with multiple known anchors, they will be assigned with distinct labels.} 
Note that the conventional GRL-based alignment algorithms need to aggregate multi-hop neighbours, ensuring the anchors can be reached to convey their carried signals. During the aggregating process, a large number of nodes in the networks indiscriminately serve as "context" for each other, making the learned representations indistinguishable for the alignment. In our proposed WL-Align, a label propagation is carried out within the individual networks, and a hashing procedure producing compressed labels is carried out across the two networks $G^s$ and $G^t$. The label of a node is iteratively updated via hashing the anchor labels and their propagating paths starting from the anchors until reaching the node (to be detailed in Section 3.2). The labels distributed over the two networks converge after three iterations. 
It is worth noting that all the nodes are eventually assigned distinct labels according to their connectivities to the known anchor. The nodes in the two networks with the same label are likely to form an anchor pair as they have essentially the same topological relationship with the given anchor. Furthermore, the distinct labels provide clues to reorganize the representation space. Incorporating the regularization of compressed labels into GRL will alleviate the ``over-smoothing'' issue when long-range anchors are encountered, resulting in more distinguishable representations for alignment tasks.

The contributions of this paper are summarized as follows:
\begin{itemize}
\item We identified that ``over-smoothing'' of node representation, encountered when embedding based alignment models aggregate long-range anchors, is one of the critical issues impairing the precision of aligning users across networks.

\item We proposed a tailored  Weisfeiler-Lehman (WL) Test for user alignment, namely across-network Weisfeiler-Lehman relabeling, to characterize the connectivities of nodes to labeled anchors, further guiding the reorganization of representation space.

\item We proposed an interactive representation learning framework that preserves second-order proximity while being regularized by the across-network Weisfeiler-Lehman labels, learning a distinguishable embedding space of nodes across networks.

\item We have evaluated the effectiveness of the proposed methodology using real-world and synthetic datasets. The experimental results obtained demonstrate that the proposed approach outperforms the state-of-the-art methods by a large margin.
\end{itemize}

\section{Related Work}
Representation learning based social network alignment can roughly be categorized into embedding-mapping and embedding-sharing methods. 
Embedding-mapping methods learn the respective node embeddings for each network first and perform subsequent mapping operations for the alignment. Using unsupervised learning, the mapping function can be designed under the assumption that there exists distribution consistency across networks. For example, REGAL~\cite{heimann2018regal} adopts xNetMF based on structural identity (structural roles) and attribute-based identity for learning node representation across networks. Node similarity can then be calculated based on the k-d tree for the alignment task. Furthermore, G-CREWE~\cite{DBLP:conf/cikm/QinSR0HK20} attempts to accelerate the matching process via a graph compression algorithm.
CONE-Align \cite{DBLP:conf/cikm/ChenHVK20} achieves the alignment via a multi-granularity strategy which preserves the consistency of the matched neighbourhood. Rather than optimizing the node-level pair-wise distance of the node embeddings, WAlign~\cite{DBLP:conf/kdd/Gao0L21}, UUIL$_{gan}$,  UUIL$_{omt}$~\cite{DBLP:conf/cikm/LiWYZZLL18} and VCNE~\cite{DBLP:conf/cikm/ChuFZB21} try to perform the alignment with respect to the entire embedding spaces. They first learn node representation via GRL algorithms such as GCN. Then, a distribution metric such as the Wasserstein distance is used to measure the discrepancy of nodes' distributions to identify potential anchors.

\begin{figure*}
	\centering
	\includegraphics[width=0.95\textwidth]{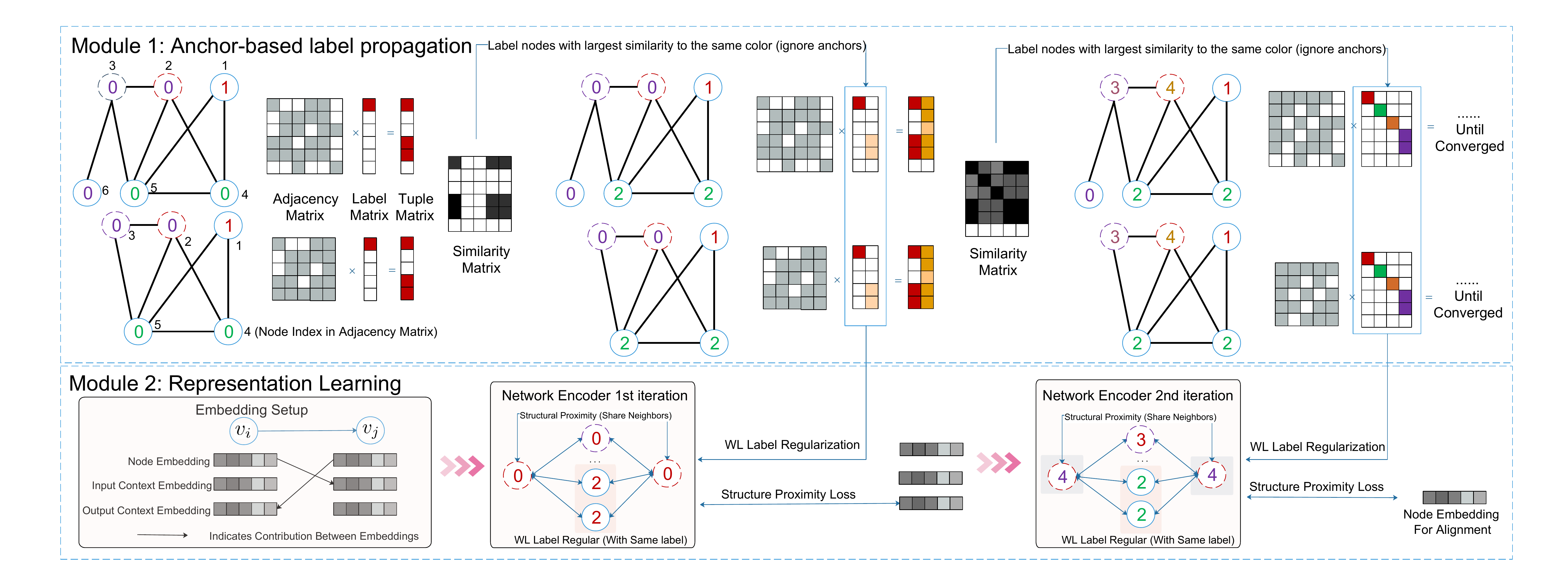}
	\caption{The Proposed Model Framework.}
	\label{modelframework}
\end{figure*}

The mapping can be achieved in a supervised fashion with the guidance of known anchors. Representative methods include PALE \cite{DBLP:conf/ijcai/ManSLJC16} which employs Multi-Layer Perceptron to map the latent space from one to another, and Deeplink \cite{DBLP:conf/infocom/0002LZTWZ18} which uses a deep neural network based dual learning process to achieve a more powerful alignment. Furthermore, SNNA \cite{DBLP:conf/aaai/LiWWYLLL19} and MSUIL \cite{DBLP:conf/cikm/LiWWLYLW19} learn the mapping function by minimizing the Wasserstein distance between the distributions of anchors across the networks. MASTER~\cite{DBLP:conf/ijcai/SuSZLQ18} and MASTER+~\cite{DBLP:journals/tkde/ZhangSSQL21} attempt to robustly reconcile multiple social networks, comprehensively exploiting attribute and structure information via an embedding approach. 

Graph neural network based models, such as MGCN \cite{DBLP:conf/kdd/ChenYS0GM20} which learns representations by leveraging convolution on both local and hypergraph network structures, exhibit their effectiveness in aligning users. iMap~\cite{DBLP:conf/cikm/XiaG021} iteratively constructs sub-graphs and adopts GCN based on the cross-entropy loss and ranking loss to perform the alignment. cM$^2$NE~\cite{DBLP:conf/kdd/XiongYP21} adopts contrastive learning on both the intra-layers (obtained by random walk and personalized pagerank) within a single network and the inter-layers across networks to learn node representations. The BANANA~\cite{DBLP:conf/ijcai/RenZZSSZG20} observes that social network alignment and behavior analysis benefit from each other and makes the first attempt to study the joint problem of social network alignment and user behavior analysis. Motivated by the inherent connection between hyperbolic geometry and social networks, 
studies~\cite{DBLP:conf/aaai/SunZZWPSY21,DBLP:conf/icdm/0008Z0WDSY20} propose to investigate user identity linkage based on hyperbolic geometry.

Embedding-sharing methods also make use of known anchors. 
They learn a unified representation space for networks with the shared embeddings of anchors. In particular, IONE~\cite{DBLP:conf/ijcai/LiuCLL16} learns the representations by preserving also the second-order follower-ship/followee-ship proximity. Several IONE-based models, such as IONE-D \cite{DBLP:journals/tkde/LiuLCL20} which takes the structural diversity into account, 
have been proposed to further boost the performance. CrossMNA \cite{DBLP:conf/www/ChuFYZHB19} leverages the cross-network information to refine shared ``inter-vectors'' and ``intra-vectors'' for the alignment.

Another line of studies related to our work is to investigate the connection between graph isomorphism testing and graph classification~\cite{DBLP:conf/iclr/XuHLJ19,DBLP:conf/aaai/0001RFHLRG19,DBLP:conf/nips/ChenVCB19}. It mainly studies the upper bounded expressive power of GNNs in distinguishing graphs and builds comparable representation models to learn continuous embedding for graph classification. In particular, 1-WL Test~\cite{DBLP:conf/iclr/XuHLJ19}, k-WL Test~\cite{DBLP:conf/aaai/0001RFHLRG19}, sigma-algebra~\cite{DBLP:conf/nips/ChenVCB19}, and ID-GNN~\cite{DBLP:conf/aaai/YouGYL21} are proposed for building the theoretical upper bound. Different from them, we take advantage of the anchor-based label propagation and hashing process of the WL Test across networks for regularizing the representation learning to make them more distinctive.

\section{Model Framework}

\subsection{Problem Definition and Overall Framework }
\textbf{Problem Definition:} Given multiple social networks $\mathcal{G}=\{ G^s, G^t\}$ to be aligned, we use superscript $s/t$ to denote the source/target network, respectively. $V^{s/t}$ denotes the set of nodes in $G^{s/t}$, where we use the term ``node(s)'' to denote the corresponding user(s) in GRL for a unified description. $A^{s/t}$ denotes the adjacency matrix of each network, structural representation learning across social networks aims to learn a function $\Phi: v_{i}^{s/t}\rightarrow \overrightarrow{u}_{i}^{s/t}$ that embeds node $v_i^{s/t}\in V^{s/t}$ into a $d$-dimensional vector $\overrightarrow{u}_{i}^{s/t}\in \mathbbm{R}^{d}$ with the structural proximity preserved. And the network alignment is to identify whether the nodes $v_i^s$ and $v_j^t$ from $G^s$ and $G^t$ form a potential anchor pair based on the learned embeddings $\overrightarrow{u}^s_i$ and $\overrightarrow{u}^t_j$.\\
\textbf{Overall Framework}: To design a representation learning framework that can take advantage of the known anchors and their connectivity to different nodes, we adopt the key ideas behind the WL Test's label propagation and the hashing procedures to guide the representation learning. We incorporate the information carried by the compressed labels, which reflect the anchors' connectivity to the corresponding nodes, into the structure-preserving objective function so that a representation space can be well-organized respecting both local and anchors-based structural information for the GRL.
Fig. \ref{modelframework} illustrates the overall framework of the proposed WL-regularized GRL. There are two modules in the learning framework. Module 1 serves to propagate the labels assigned to the anchors and ``compress'' (via hashing) the set of labels propagated to each node. Module 2 is the GRL module, which utilizes the compressed labels to learn representations along with anchor-based label propagation while maintaining ``second-order'' structural proximity among the neighbouring nodes.

\subsection{Across-network Weisfeiler-Lehman Relabeling}
\label{sect:WL-Align}
In this subsection, we introduce module 1 which carries out anchor-based label propagation and label compressing steps across the networks. It is similar to the Weisfeiler-Lehman (WL) Test, but tailored for the social network alignment task.  
We first initialize the label of each anchor pair to a distinct number while setting the label of non-anchor nodes to 0. Then, the labels are propagated over the networks. For each propagation step, each non-anchor node will be assigned a tuple that contains the label of the node itself and a multiset with its first-order neighbours' labels.\footnote{A multiset is a set in which elements may appear multiple times and the order is not important.}

Mathematically, we represent the label of the $i$-th node $v_i^{s/t}$ using a one-hot vector, denoted as $\overrightarrow{wl}_{i}^{s/t}\in \{0,1\}^{|C_a|}$ where $\sum wl_i=1$ and $|C_a|$ is the size of the label set. For the node with label 0, $\overrightarrow{wl}_{i}^{s/t}$ will be set as a zero vector, indicating that there is no external information associated with non-anchor nodes for the alignment. Initially, $|C_a|$ is equal to the size of the anchor set $V_a$, which denotes that each anchor pair has its own distinct one-hot vector and is linearly independent of the others.
Then, we stack up $\overrightarrow{wl}_{i}^{s/t}$ for all the nodes in $G^s$ and $G^t$ to form the label matrices $WL^{s}$ and $WL^{t}$, respectively.
Each iteration of the label propagation becomes:

\begin{equation}
    \label{wl_propagate}
    {WL}_{tp}^{s/t}=\overline{A}^{s/t}\times{WL^{s/t}}
\end{equation}
where $\overline{A}^{s/t}=A^{s/t}+I$ ensures that the label of the node itself will be included in the updated tuple of labels. ${WL}_{tp}$ is referred to as the \textit{tuple matrix} in the sequel.

\begin{figure}
	\centering
	\includegraphics[width=0.45\textwidth]{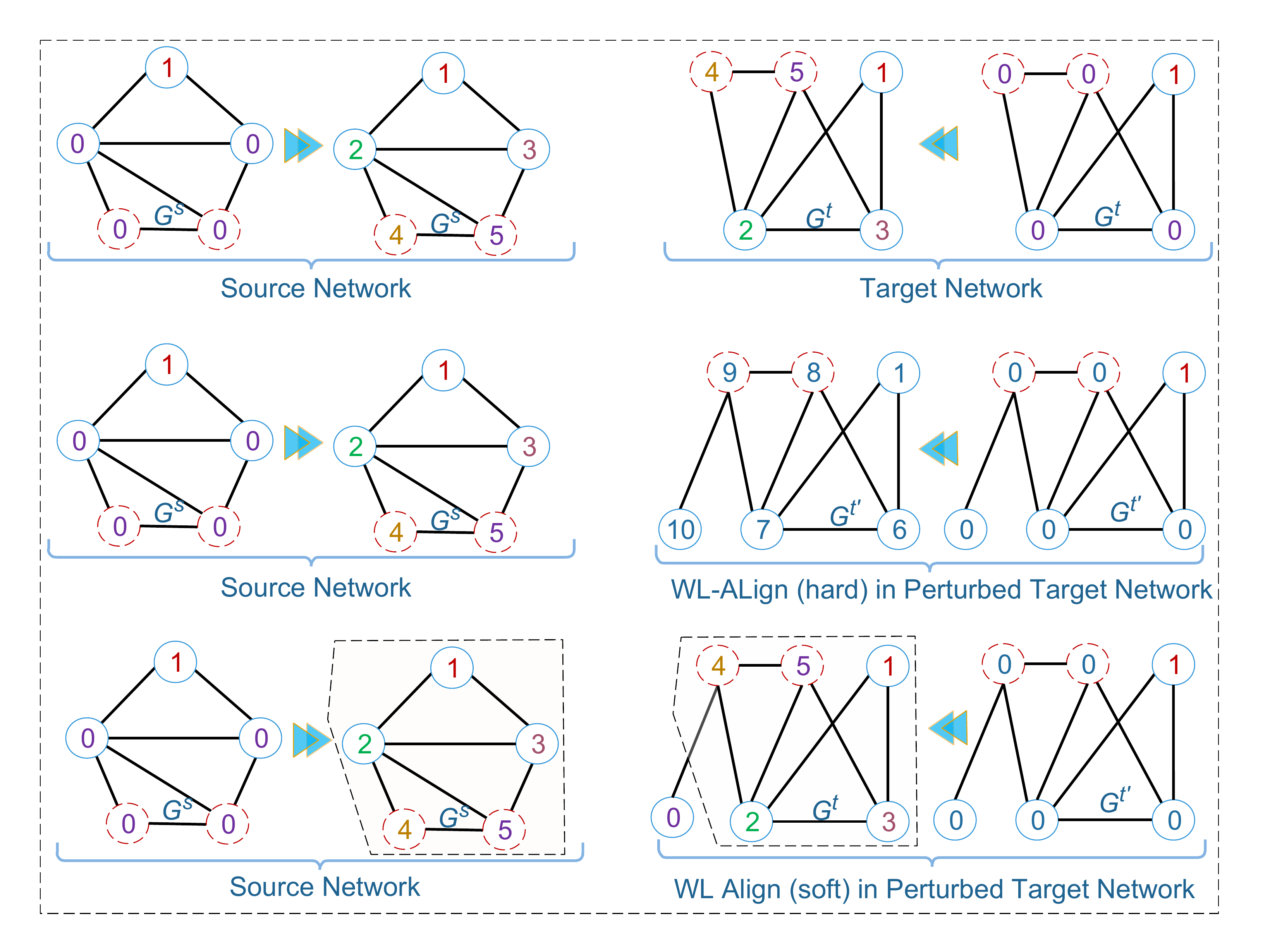}
	\caption{An illustrated example showing the robustness of the anchor-based label propagation in WL-Align. See Appendix \ref{appendix:WLHard} for the hash rules used in perturbed networks.}
	\label{ToySimWL}
\end{figure}

Then, we relabel nodes by first ignoring the elements of 0 in the tuple and then hashing the tuples via an injective hash function. We ignore the elements of 0 in the tuple as there is no identifiable information associated with the non-anchor nodes for the alignment. An integer based auto-incremental function is used to compress labels in the tuples of each node (see \textit{Hash Rules} in Fig. \ref{ToyExamOne}). We can then repeat the iterative relabeling process until convergence (as illustrated in the first row of Fig. \ref{ToySimWL}).

In practice, the networks to be aligned will not be structurally identical.
We notice that the hashing procedures based on identical tuples in the WL Test (named \textit{WL-Align (hard)} in this paper) do not work well for two graphs with only a slight difference. As shown in the second row of Fig. \ref{ToySimWL}, the network $G^t$ is perturbed by implanting only one additional node with label 0 to it. When the relabeling process is converged, the nodes' labels deviate a lot from what they would be without the perturbation.
To mitigate the sensitivity to the arbitrariness of network structure, we proposed \textit{WL-Align (soft)}, which relaxes the hash function by computing the largest similarity between tuples across networks. 

After each round of propagation, we perform a one-hot hash encoding to re-label nodes according to their newly generated tuples -- that is the row vectors on the left-hand side of Eq. \eqref{wl_propagate}. However, it is hard for nodes in the two different networks to have exactly the same tuple.
So instead of using the injective/perfect hash function in \textit{WL-Align (hard)} to indicate alignment (as shown in the second row of Fig. \ref{ToySimWL}), \textit{WL-Align (soft)} utilizes an imperfect hash function to assign labels based on the similarity of the tuples across networks. We compute the similarity between the two tuple matrices as follows:
\begin{gather}
    \label{wl_similarity}
    WL^{sim}=\frac{{WL}_{tp}^{s}}{||{WL}_{tp}^{s}||_2}\times\frac{{{WL}_{tp}^{t}}^T}{||{{WL}_{tp}^{t}}^T||_2}
\end{gather}
where ${WL}_{tp}^{s/t}$ is normalized by the row-wise L2 norm to mitigate the influence of vastly different degrees of nodes. Besides, the use of the normalized $WL^{s/t}$ can guarantee the nodes with identical tuples (the same vector after one propagation in Eq. \eqref{wl_propagate}) to give the highest similarity. This is also consistent with the WL Test where nodes should be hashed to the same label if they have the same tuple.

We then assign $v_i^s$ and $v_j^t$ with the same label on the condition that $v_i^s$ and $v_j^t$ are the most similar nodes to each other across the networks, which is formulated as Eq. \eqref{relabel}:
\begin{gather}
    \nonumber
    WL_{ij}^{sim} \geq \forall WL_{ik}^{sim}, j\neq k \ \\
    \nonumber
    and \ WL_{ij}^{sim} \geq \forall WL_{kj}^{sim}, i\neq k.\\
    and \ WL_{ij}^{sim} > 0.
\label{relabel}
\end{gather}
The indices of users who should be labeled are obtained using the intersection of two sets which look up the maximum values of each row and column of the matrix $WL^{sim}$. For instance, if $WL^{sim}$ is computed according to Eq.\eqref{wl_similarity} gives:
\begin{gather}
    \label{wl_example}
    WL^{sim}= \left\{ \begin{matrix} 0 & 1 & 0.4 \\ 1 & 0.6 & 0.3\\ 0 & 0 & 1 \\ \end{matrix} \right\}.
\end{gather}
$WL_{01}^{sim}$, $WL_{10}^{sim}$, and $WL_{22}^{sim}$ are the maximum values in their respective rows and columns. $v_0^s$ and $v_1^t$ will be relabeled with the same one-hot compressed label, and so on for $v_1^s$ and $v_0^t$, $v_2^s$ and $v_2^t$.

We perform the aforementioned process iteratively until the size of the label set, i.e., $|C_a|$ does not change anymore. 
We anticipate that this anchor-based label propagation strategy is able to explore the distinguishable connectivities of nodes to long-range anchors and can regularize the subsequent graph representation learning (as shown in the third row of Fig. \ref{ToySimWL}). 

\subsection{Regularized Representation Learning}
As the node labels derived from the anchor-based propagation step can reveal the anchors' connectivity to the nodes, Module 2 makes use of them to reduce the chance of inaccurate matching which usually happens in the over-smoothing representation space resulting from aggregating long-range nodes.
We propose the following objectives to guide the representation learning:\\

\noindent
\textbf{Objective 1: } To achieve the fact that nodes with the same compressed labels across the social networks should be close and nodes with different labels should be far apart in the embedding space, we define a loss function $\mathcal{L}_{label}$, given as Eq. \eqref{loss_label}:
\begin{gather}
    \nonumber
    \mathcal{L}_{label}\left(v_{i}^s, v_{j}^{t}\right)  = (2 L_{ij} - 1) \left[cos\left({\overrightarrow{u}_i^{s}}^T \overrightarrow{u}_j^{t}\right)\right] \\
    L_{ij} =  \begin{cases}1, & {label}\left(v_{i}^{s}\right) ={label}\left(v_{j}^{t}\right) \\ 0, & {label}\left(v_{i}^{s}\right) \not= {label}\left(v_{j}^{t}\right)\end{cases}
    \label{loss_label}
\end{gather}
where $L_{ij}$ denotes whether $v_i^{s}$ and $v_j^{t}$ have the same compressed label,   $\overrightarrow{u}_i^{s}$ and $\overrightarrow{u}_j^{t}$ are the corresponding embeddings of $v_i^{s}$ and $v_j^{t}$. And the cosine function $cos(\cdot)$ is leveraged to learn the normalized similarity between embeddings of users. 

$\mathcal{L}_{label}$ helps preserve the proximity of nodes with the same label across networks by considering them as ``positive'' pairs in the context of sample-wise learning, which can ensure that the nodes across networks with similar connectivity with reference to the anchors are close in the embedding space. When there are multiple nodes associated with the same label within a single network, they will be selected to form positive pairs with the node having the same label in another network. Thus, the label proximity within each network can also be preserved via $\mathcal{L}_{label}$. Since the nodes with the same label within each network cannot be differentiated by the conventional structural representation learning, $\mathcal{L}_{label}$ drives them apart from others with the help of negative pairs, reducing the chance of inaccurate matching in the embedding space.

In addition to preserving the proximity between nodes with similar anchor diffused labels, we also enforce preserving second-order structural proximity in our framework to fully utilize comprehensive structure patterns of labeled and unlabeled nodes in the representation learning.\\
\noindent
\textbf{Objective 2: } To allow nodes with a similar structural context within each network to be close and the anchor pairs across networks to have identical representations for learning a unified embedding space,
To form objective 2, we define the corresponding loss function $\mathcal{L}_{context}$ as shown in Eq. \eqref{loss_context}:
\begin{align}
    \nonumber
    \mathcal{L}_{context}\left(v_{i}^{s/t}, v_{j}^{s/t}\right)= 
    C_{ij} \log \sigma\left({\overrightarrow{u}_{i}^{s/t}}^{T} \overrightarrow{u}_{j}^{'s/t}\right) +\\
    \nonumber
    C_{ij} \log \sigma\left({\overrightarrow{u}_{i}^{''s/t}}^{T} \overrightarrow{u}_{j}^{s/t}\right)+\\
    \nonumber
     \left(1-C_{ij}\right) \log \sigma\left(-{\overrightarrow{u}_{i}^{s/t}}^{T} \overrightarrow{u}_{j}^{'s/t}\right)+\\
    \nonumber
    \left(1-C_{ij}\right) \log \sigma\left(-{\overrightarrow{u}_{i}^{''s/t}}^{T} \overrightarrow{u}_{j}^{s/t}\right)
        \label{loss_context}
\end{align}
\begin{align}
    C_{ij}= \begin{cases}1, & v_{j} \in \operatorname{context} \left(v_{i}\right) \\ 0, & otherwise \end{cases}
\end{align}
in which $\mathcal{L}_{context}$ is defined on a directed edge from $v_i$ to $v_j$, as shown in the representation module of Fig. \ref{modelframework}. Similar to IONE~\cite{DBLP:conf/ijcai/LiuCLL16}, we define $\overrightarrow{u}_{i}^{s/t}$, $\overrightarrow{u}_{i}^{'s/t}$ and $\overrightarrow{u}_{i}^{''s/t}$ as the corresponding node representation, ``input context'' representation and ``output context'' representation w.r.t. node $v_i$. As illustrated in the bottom left of Fig. \ref{modelframework}, given an edge $v_i\rightarrow v_j$, the embedding of $v_i$ will contribute the ``input context'' embedding of $v_j$ while the embedding of $v_j$ will contribute ``output context'' embedding of $v_i$. Then, by optimizing the $\mathcal{L}_{context}$, all three types of embeddings are updated simultaneously according to the follower-ship and followee-ship. And the follower-ship and followee-ship collaboratively define one’s unique social figure in the social network. The undirected social network can be transformed into a bi-directed network to adopt this learning schema. $C_{ij}$ denotes whether $v_j$ is the context node (defined as a 1-hop neighbour in this paper) of $v_i$, and $\sigma$ denotes the sigmoid function.

The node representations in the multiple networks can then be learned by optimizing the combined objective function, shown in Eq. \eqref{finalloss}:
\begin{align}
\label{finalloss}
\mathcal{L}= \mathcal{L}_{label} + \mathcal{L}_{context}.  
\end{align}
Note that the node representations $\overrightarrow{u}_{i}^{s/t}$ influenced by optimizing $\mathcal{L}_{label}$ can contribute to their corresponding input/output representations through the contribution paths defined in $\mathcal{L}_{context}$. In this way, nodes sharing contextual nodes with the same labels will also be made close in the representation space. Therefore, $\mathcal{L}$ can take advantage of both the $\mathcal{L}_{label}$ and $\mathcal{L}_{context}$ via the shared parameters $\overrightarrow{u}_{i}^{s/t}$. We use the Adam optimizer \cite{DBLP:journals/corr/KingmaB14} to learn the node representations of the two networks. The overall algorithm is shown in Algorithm \ref{algo}. To map nodes across different social networks, we compute the cosine similarity between the representation of one node in network $G^s$ and another in network $G^t$. For each node $v_{i}^s$ in network $G^s$, we can determine whether $v_{j}^t$ in network $G^t$ is a potential anchor candidate based on the ranking of the defined relevance.
\begin{algorithm}
	\caption{Weisfeiler-Lehman regularized representation learning across networks}\label{algo}
	\algorithmicrequire ~Two networks $G^s$ and $G^t$, a set of anchors $V_a$, the number of negative samples for $\mathcal{L}_{label}$ $K_L$, the number of negative samples for $\mathcal{L}_{context}$ $K_C$, the training batch set $T$, number of embedding updating epoch $E$ \\
	\algorithmicensure ~The set of estimated parameters $\Theta=\{\overrightarrow{u_{i}},\overrightarrow{u_{j}},\overrightarrow{u_{i}}'',\overrightarrow{u_{j}}'\}$
	\begin{algorithmic}[1]
		\Procedure{Learning}{$G^X$,$G^Y$, $V_a$, $K$}
		\State Initialize $\Theta=\{\overrightarrow{u_{i}},\overrightarrow{u_{j}},\overrightarrow{u_{i}}'',\overrightarrow{u_{j}}'\}$
		\State Initialize $WL^s$ and $WL^t$ according to $C_a$ (using anchors $V_a$).
		\Repeat
		\State Calculate $WL^{sim}$ according to Eq. \eqref{wl_similarity}.
		\State Compress labels of nodes to obtain $C_a$ based on rules of Eq. \eqref{relabel}.
		\State  Update $WL^s$ and $WL^t$ according to $C_a$.
		\State Constructing training batch set T via sampling node pair $(v_i^{s}$, $v_j^{t})$ along with $K_L$ negative pairs across the networks, and $(v_{i}^{s/t}$,$v_{j}^{s/t})$ along with $K_C$ negative pairs within each of the networks.
		\For{$i=0;i<E;i=i+1$}
		\For{each instance in $T$}
		\State Update $\overrightarrow{u_{i}}, \overrightarrow{u_{j}}, \overrightarrow{u_{j}}', \overrightarrow{u_{i}}'' $ according to \eqref{loss_label} and \eqref{loss_context} based on the Adam optimizer.
		\EndFor
		\EndFor
		\Until Model Convergence
		\State \textbf{return} ${\Theta}$
		\EndProcedure
	\end{algorithmic}
\end{algorithm}

\section{Experiment and Analysis}

For performance evaluation, we applied the proposed framework to a synthetic dataset of two correlated  Erd{ő}s-R{é}nyi (E-R) graphs and three widely adopted real-world datasets, namely Twitter-Foursquare~\cite{DBLP:conf/ijcai/LiuCLL16,r6,DBLP:conf/ijcai/ZhangY15}, ACM-DBLP~\cite{Nettrans,NeXtAlign} and Phone-Email~\cite{Nettrans, DBLP:journals/tkdd/ZhangTTXF20}, and compared its performance with the state-of-the-art user alignment methods. Table \ref{dataset} lists the statistics of the real-world datasets.

The synthetic dataset is constructed for evaluating the robustness of the proposed \textit{WL-Align (soft)} in the across-network Weisfeiler-Lehman relabeling module. We first generate an E-R random graph \cite{erdHos1960evolution} of 1,000 nodes, where nodes are randomly connected with A probability of 0.01, yielding 5,064 edges. Then, we augment the E-R graph by adding random nodes and/or connections to produce augmented graphs. 
We form pairs of the augmented graphs, each with the same percentage of nodes or edges added, to constitute the synthetic dataset.

For the Twitter-Foursquare dataset, the ground truth is obtained by finding users who provide their twitter accounts in Foursquare profiles. For the ACM-DBLP dataset, the co-author networks are formed by connecting edges between users who are co-authors of at least one paper in four areas (DM, ML, DB, and IR). The same authors in both ACM and DBLP are considered as the ground truth. For the Phone-Email dataset, networks are constructed by the interaction of a group of people through different communication channels, including phone and email. The pair of networks have 1,000 persons in common who serve as the ground truth.

\begin{table} \centering 
	\caption{Statistics of the datasets used for evaluation.}
	\label{dataset}
	\begin{tabular}{c||c|c|c}
		\hline
		\hline
		Networks&\#Users&\#Relations&\#Anchors\\
		\hline
		\hline
		Twitter&5,220&164,919&\multirow{2}{*}{1,609}\\
		\cline{2-3}
		Foursquare&5,315&76,972&\\
		\hline
		ACM&9,872&39,561&\multirow{2}{*}{6,325}\\
		\cline{2-3}
		DBLP&9,916&44,808&\\
		\hline
		Phone&1000&41,191&\multirow{2}{*}{1000}\\
		\cline{2-3}
		Email&1003&4,627&\\
		\hline
	\end{tabular}
\end{table}

\subsection{Baseline Models and Evaluation Metric}
Since the proposed WL-Align runs in a semi-supervised manner, we compare it with five state-of-the-art (semi-) supervised methods, including shallow GRL based \textbf{IONE} \cite{DBLP:conf/ijcai/LiuCLL16} and GNN based \textbf{DEEPLINK} \cite{DBLP:conf/infocom/0002LZTWZ18}, \textbf{DANA} \cite{DANA}, \textbf{NetTrans} \cite{Nettrans} and \textbf{NeXtAlign} \cite{NeXtAlign} \footnote{The codes used in the experiments can be found at \url{https://github.com/ColaLL/IONE}, \url{https://github.com/KDD-HIEPT/DeepLink}, \url{https://github.com/xhhszc/DANA}, \url{https://github.com/sizhang92/NetTrans-KDD20}, \url{https://github.com/sizhang92/NextAlign-KDD21}.}.

\begin{itemize}
\item \textbf{IONE} \cite{DBLP:conf/ijcai/LiuCLL16} adopts a shallow graph neural network to learn node representations, in which follower and followee relationships are explicitly represented as input context and output context vectors. By preserving the second-order proximity, IONE learns a unified latent space for the alignment. It is to be noted that IONE can also be considered a reduced version of our proposed WL-Align by using only Eq.\eqref{loss_context} to accomplish the alignment.

\item \textbf{DEEPLINK} \cite{DBLP:conf/infocom/0002LZTWZ18} applies deep neural network based mapping and dual learning to social network alignment. The random walk and skip-gram algorithms are utilized for learning the initial representations of networks, and then the mapping function is learned using two MultiLayer Perceptrons.

\item \textbf{DANA} \cite{DANA} employs a domain adversarial mechanism to obtain domain-invariant representations of different networks for the alignment task.

\item \textbf{NetTrans} \cite{Nettrans} applies an end-to-end model that learns a composition of nonlinear functions to transform one network to another, where two poolings operations, TransPool and TransUnPool, are proposed for encoding and decoding networks during the alignment process.

\item \textbf{NeXtAlign} \cite{NeXtAlign} studies the disparity issue in network alignment. To achieve a good trade-off between alignment consistency and alignment disparity, it proposes a special graph convolutional network RelGCN-U and alignment scoring function for the representation learning.

\end{itemize}

\noindent
\textbf{Evaluation Metric:} We adopt $Precision@N$  \cite{DBLP:conf/ijcai/LiuCLL16,DBLP:conf/infocom/0002LZTWZ18,DANA,NeXtAlign} as the evaluation metric, defined as:
\begin{eqnarray}\label{p_at_t}
Precision@N=\frac{|CorrUser@N|^X+|CorrUser@N|^Y}{|UnMappedAnchors|\times2}
\end{eqnarray}
where $|CorrUser@N|$ is the number of unmapped anchor nodes with their corresponding nodes found among the top-$N$ neighbours in the embedded space. $|UnMappedAnchors|$ is the total number of all unmapped anchor nodes. 

\noindent
\textbf{WL-Align Implementation Details: }The proposed model is implemented in Pytorch. We use one Nvidia GTX 2080ti as GPU. We use the Adam optimizer with a learning rate of 0.05 to optimize the model. We set identical hyperparameters for three datasets. Where the batch size is set to 1000. And the numbers of negative samples $K_L$ and $K_C$ (shown in Algorithm \ref{algo}) are set to 1 and 20, respectively. Iterations of representation learning module $E$ is set to 50. The dimension of user representation is set to 128.

\begin{figure}
	\centering
	\subfigure[Perturbation by adding nodes]{
	\label{synexpNode}
		\includegraphics[width=0.46\columnwidth]{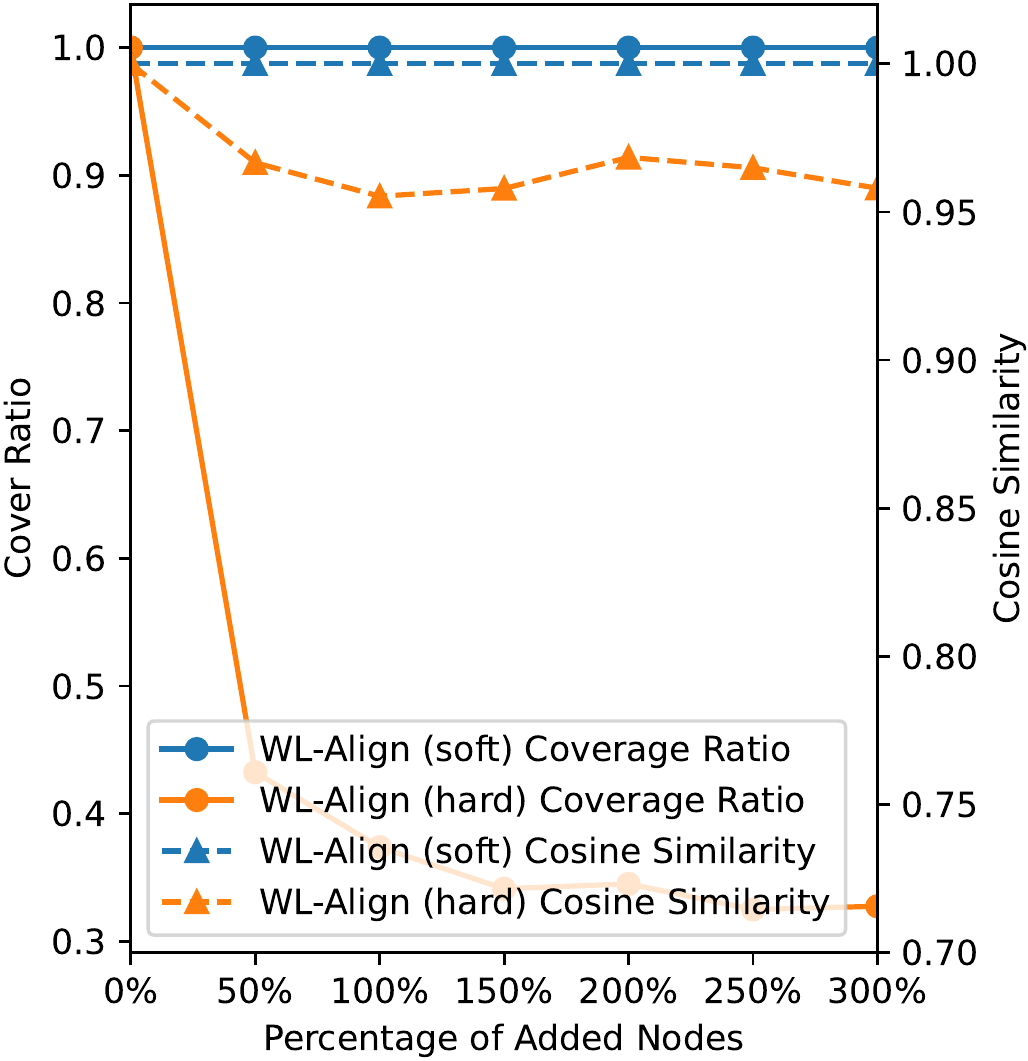}
	}
	\subfigure[Perturbation by adding edges] { \label{synexpEdge}
		\includegraphics[width=0.45\columnwidth]{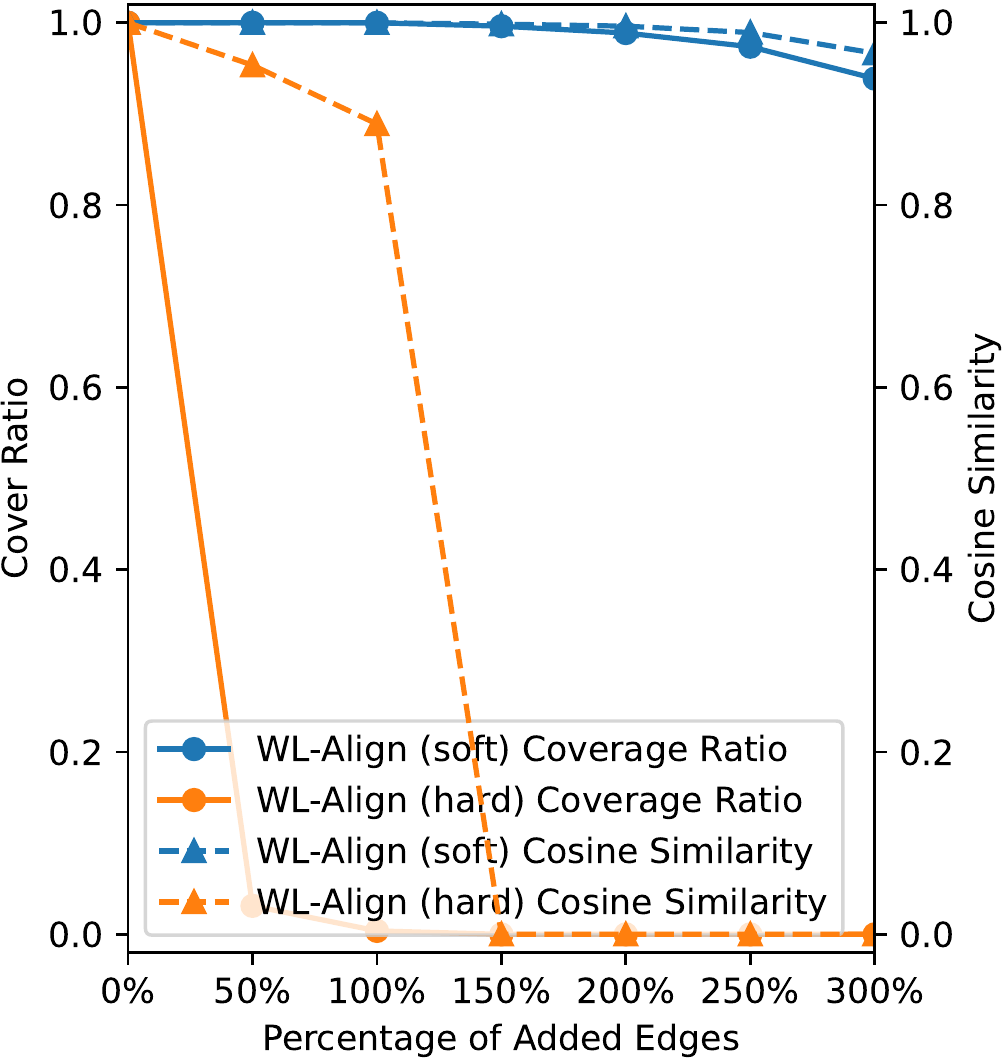}
	}
	\caption{Performance of WL-Align given different degrees of perturbation in the network pair. 
	Note that we shift the axis of ``cosine similarity'' for better illustration.}
	\label{SynExp}
\end{figure}

\begin{figure*} \centering
	\subfigure[Twitter-Foursquare Dataset]{
	\label{50PERCENTTF}
		\includegraphics[width=0.3\textwidth]{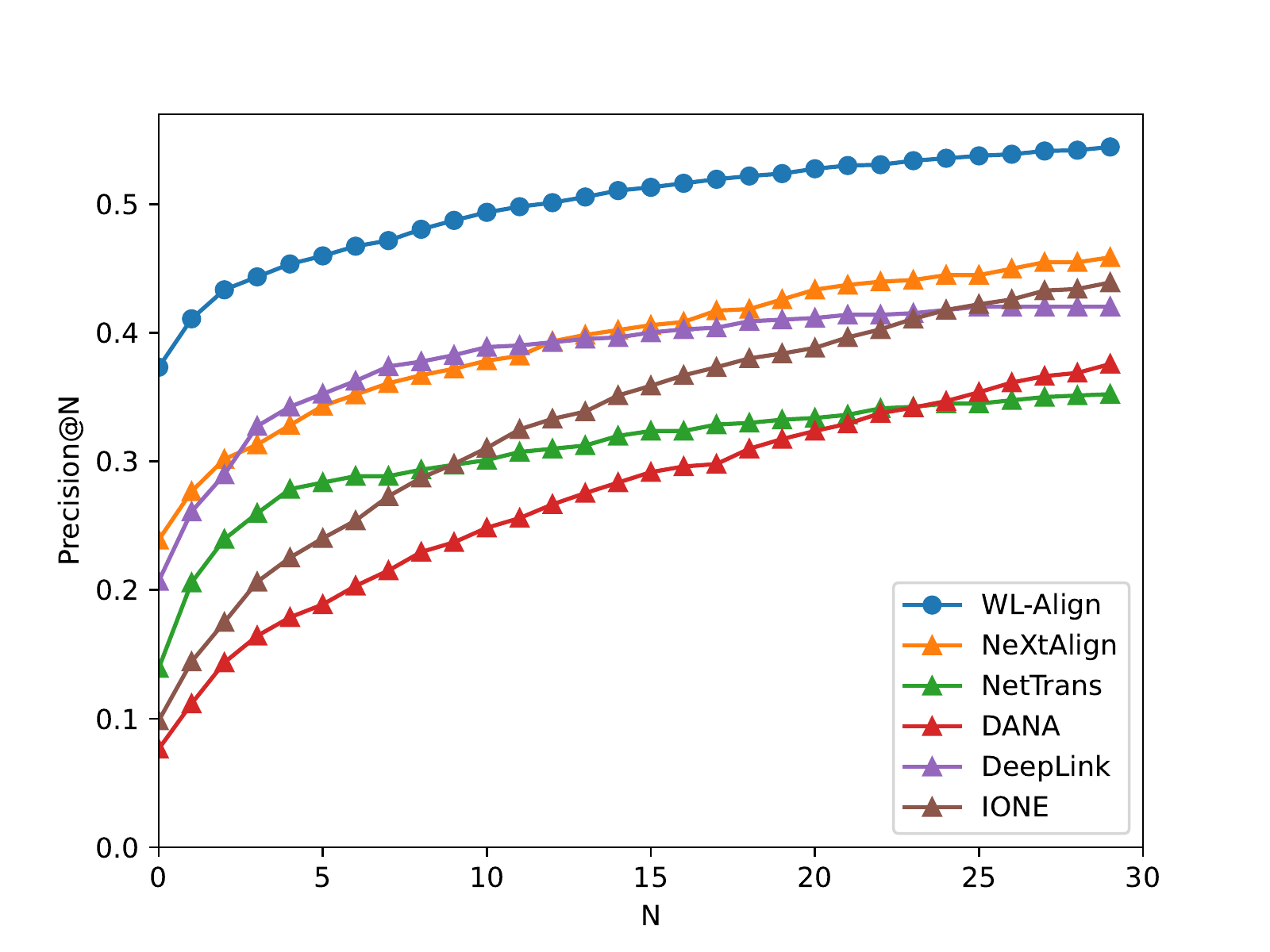}
	}
	\subfigure[ACM-DBLP Dataset] { \label{50PERCENTAD}
		\includegraphics[width=0.3\textwidth]{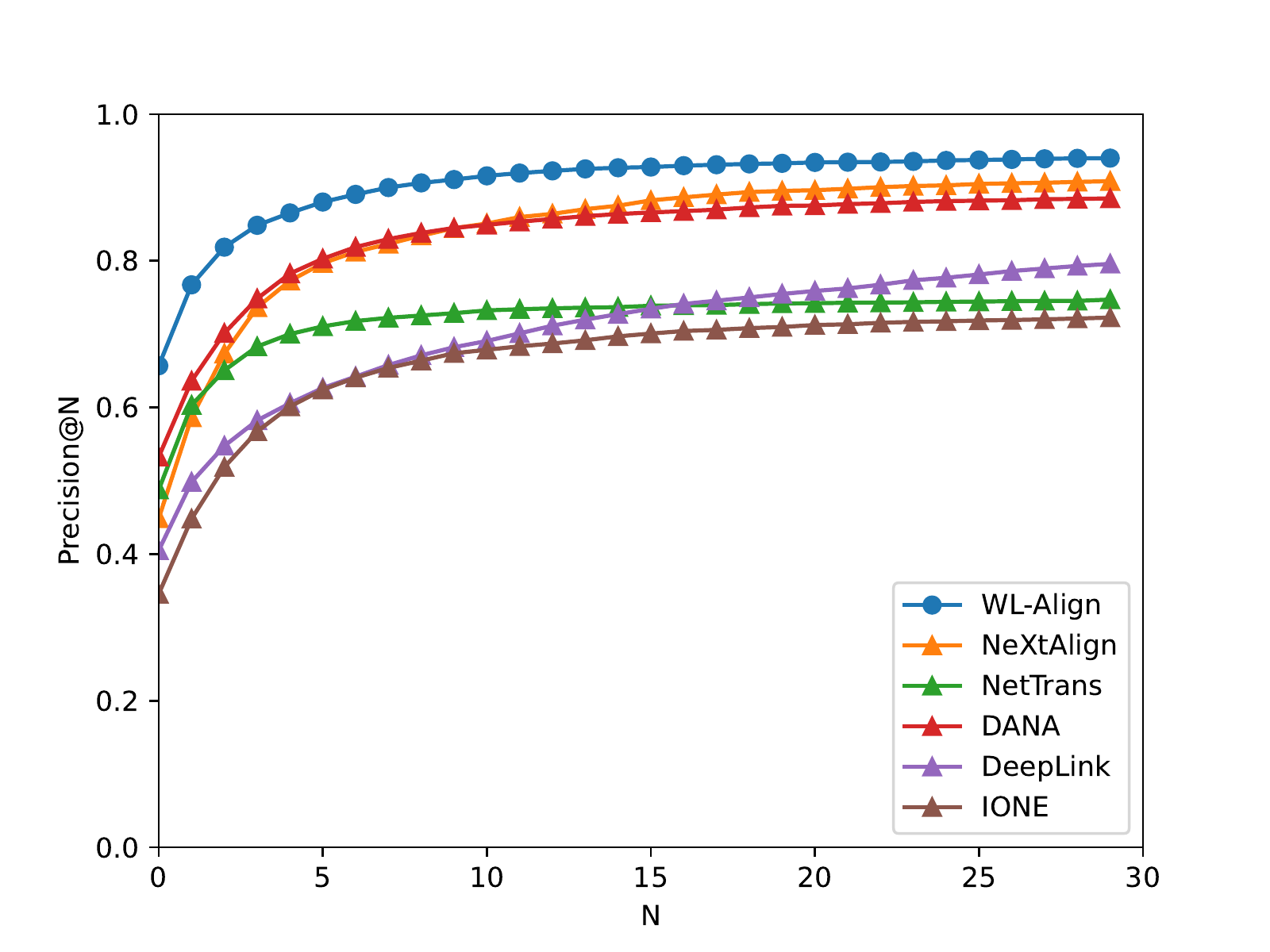}
	}
	\subfigure[Phone-Email Dataset]{
	\label{50PERCENTPE}
		\includegraphics[width=0.3\textwidth]{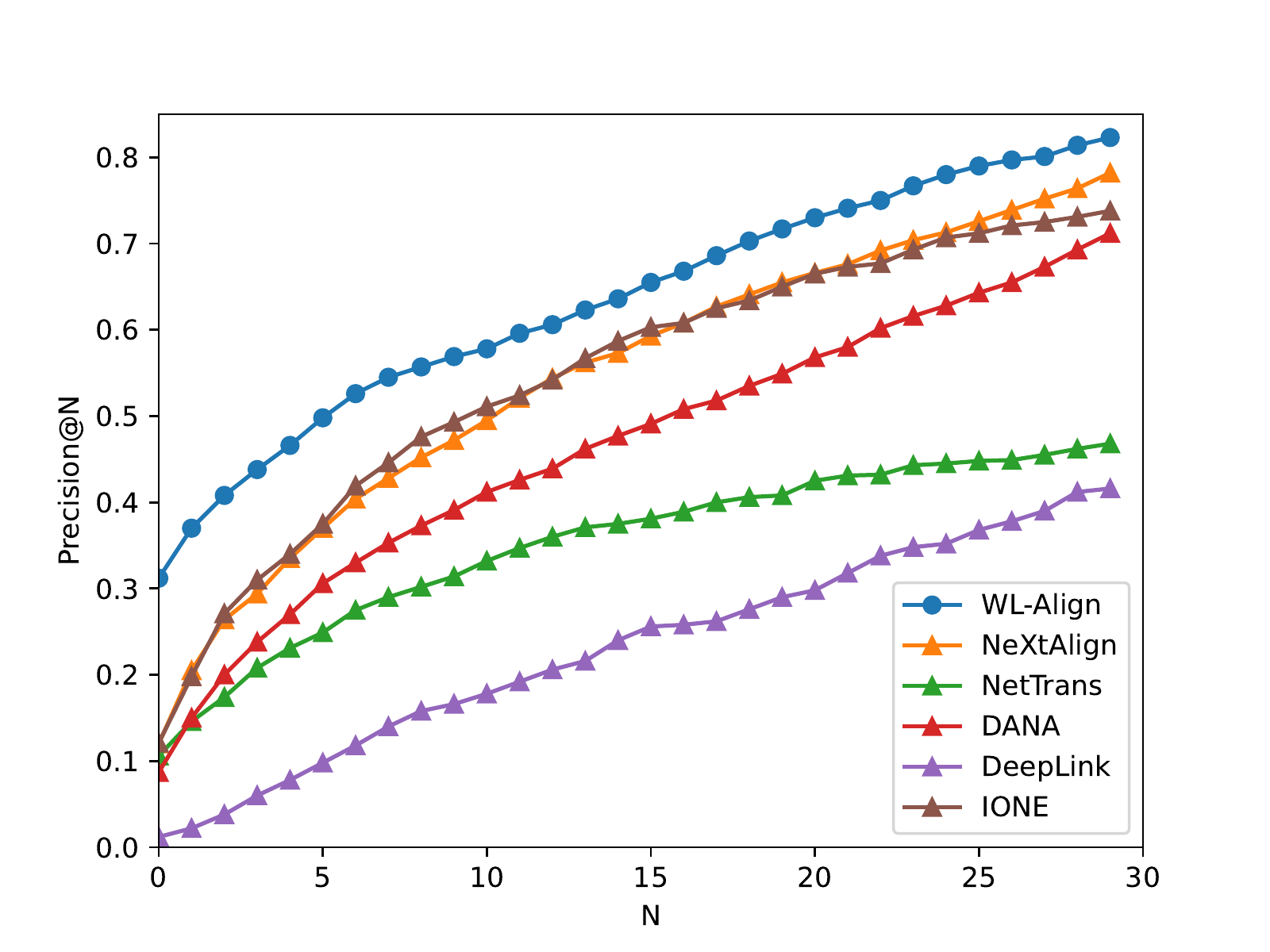}
	}
	\caption{Precision@1-30 Performance under Training Ratio 50\%.}
	\label{TrainingRatio50}
\end{figure*}

\begin{figure*} \centering
    \subfigure[Twitter-Foursquare Dataset]{
	\label{TFDifRatio}
		\includegraphics[width=0.3\textwidth]{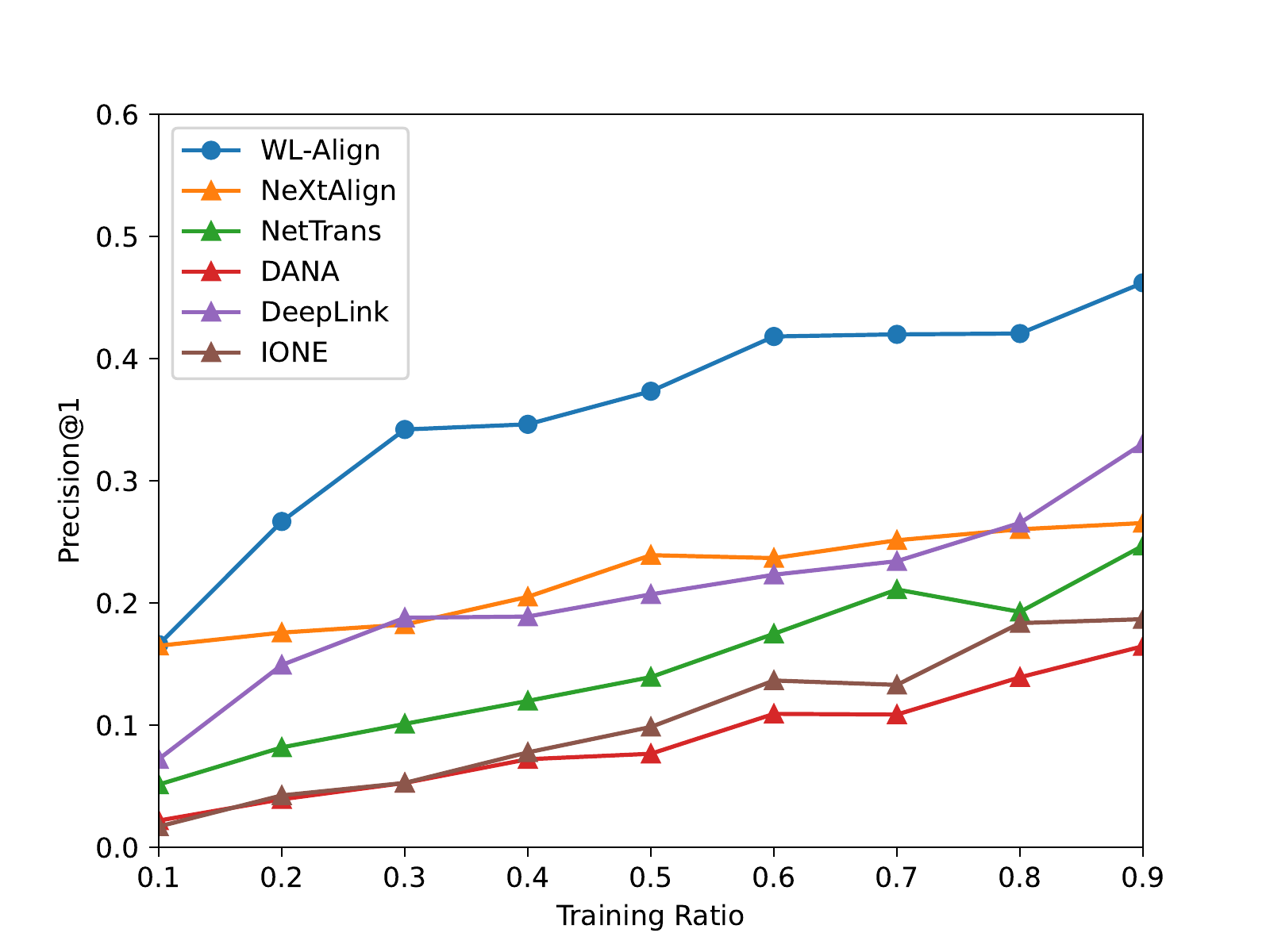}
	}
	\subfigure[ACM-DBLP Dataset] { \label{ADDifRatio}
		\includegraphics[width=0.3\textwidth]{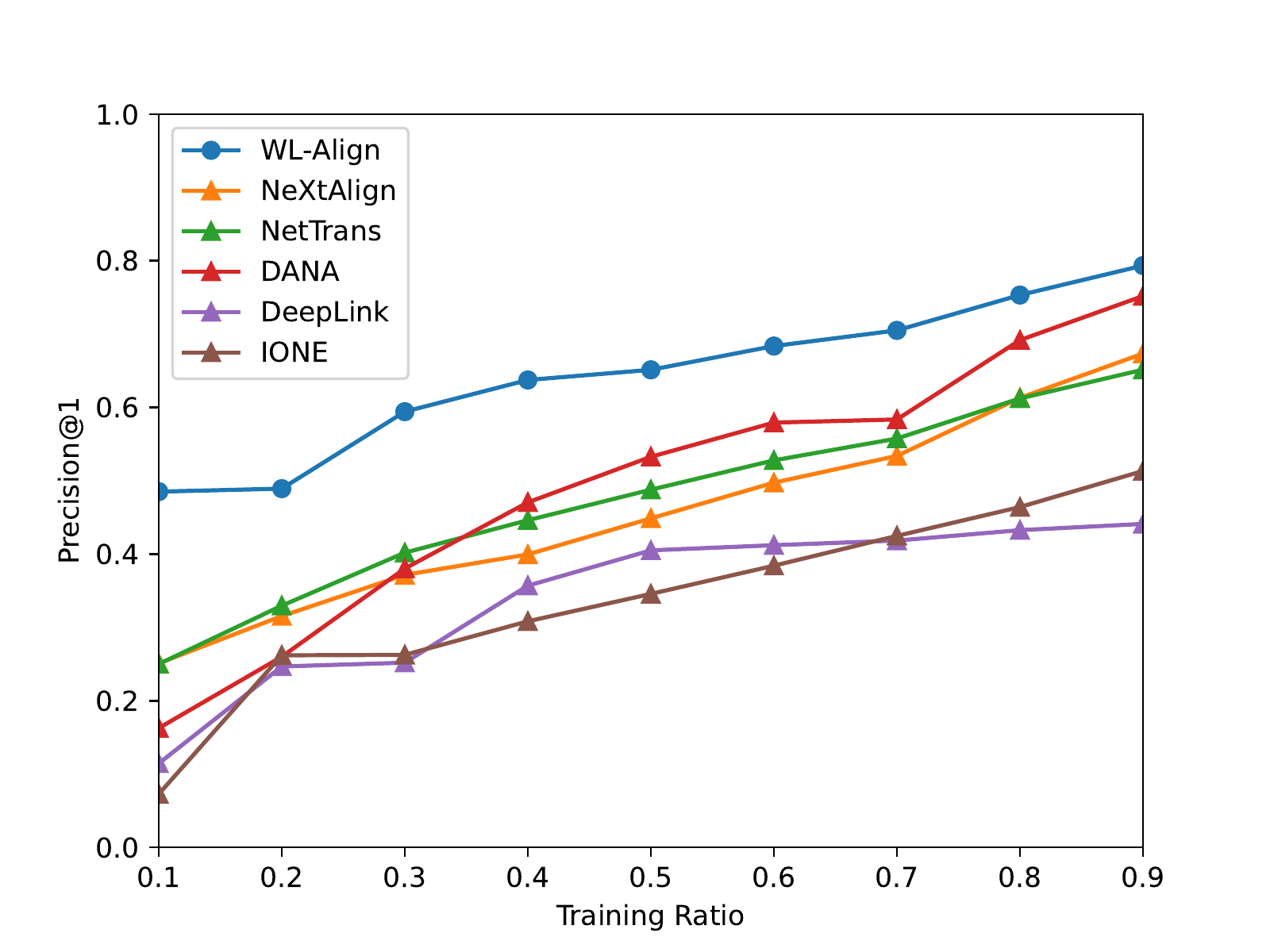}
	}
	\subfigure[Phone-Email Dataset]{
	\label{PEDifRatio}
		\includegraphics[width=0.3\textwidth]{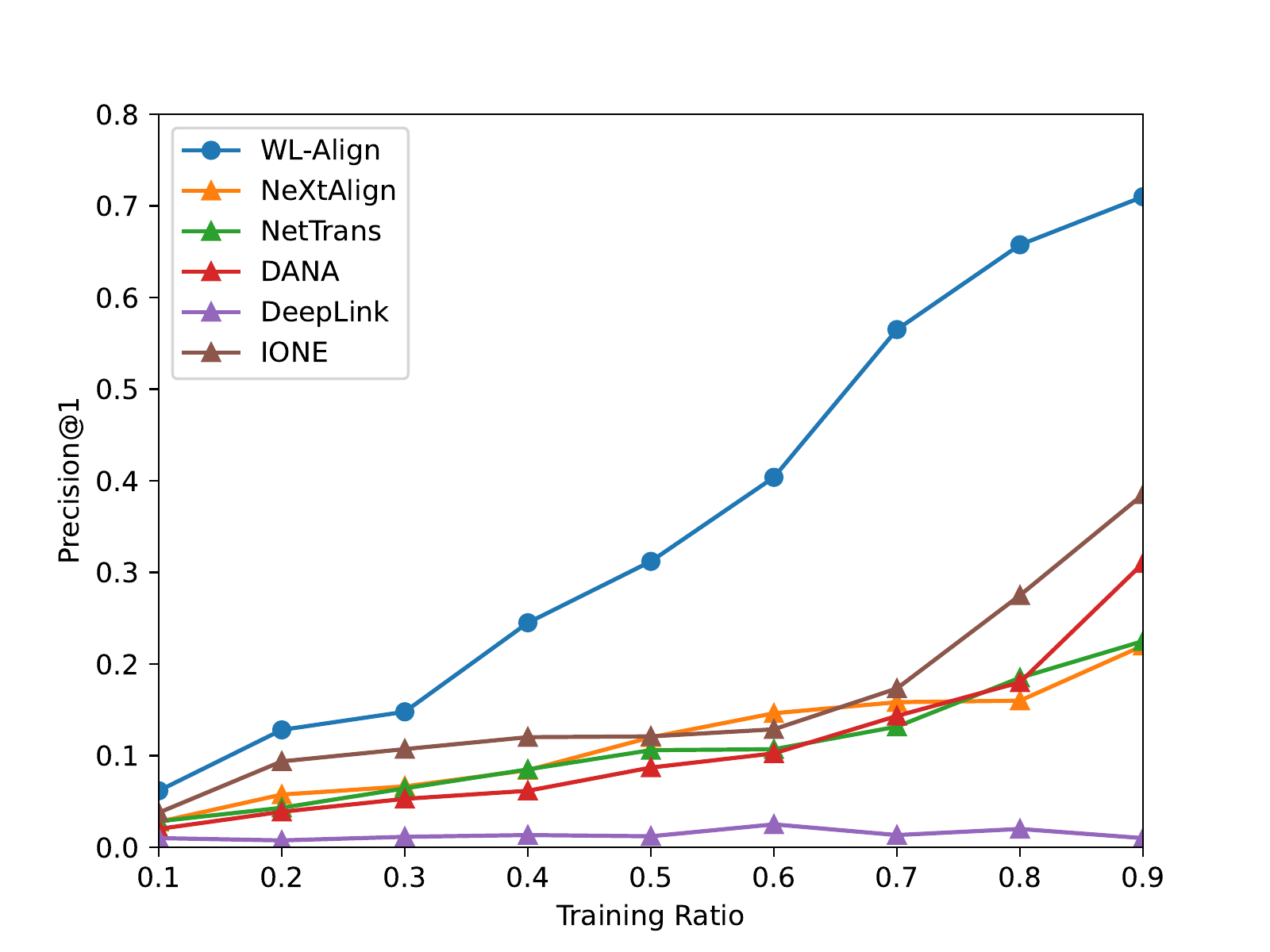}
	}
	\caption{Performance under Different Training Ratios.}
	\label{DifferentTrainingRatio}
\end{figure*}

\begin{figure*} \centering
	\subfigure[Twitter-Foursquare Dataset]{
	\label{motivationft}
		\includegraphics[width=0.3\textwidth]{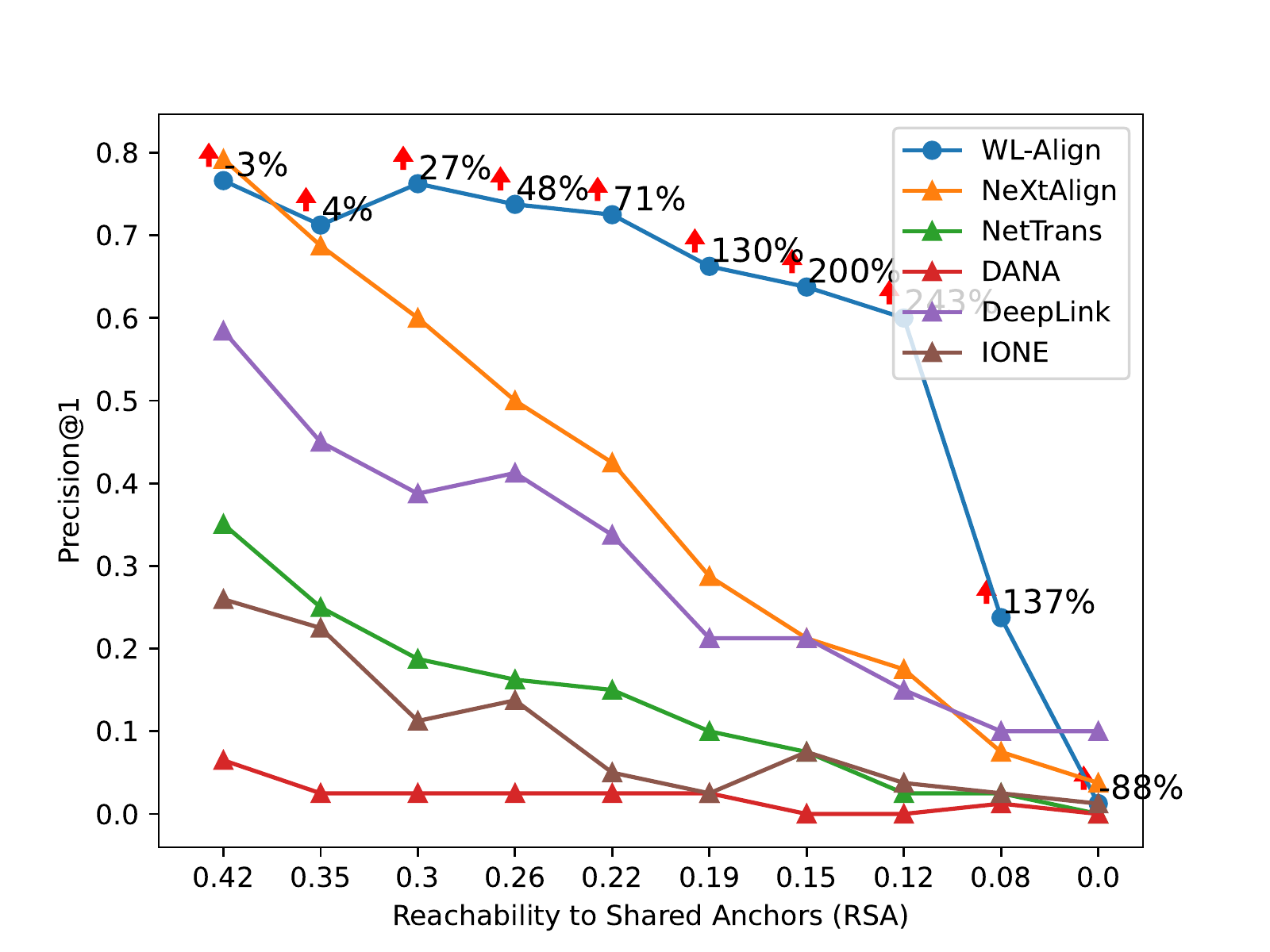}
	}
	\subfigure[ACM-DBLP Dataset] { \label{motivationad}
  \includegraphics[width=0.3\textwidth]{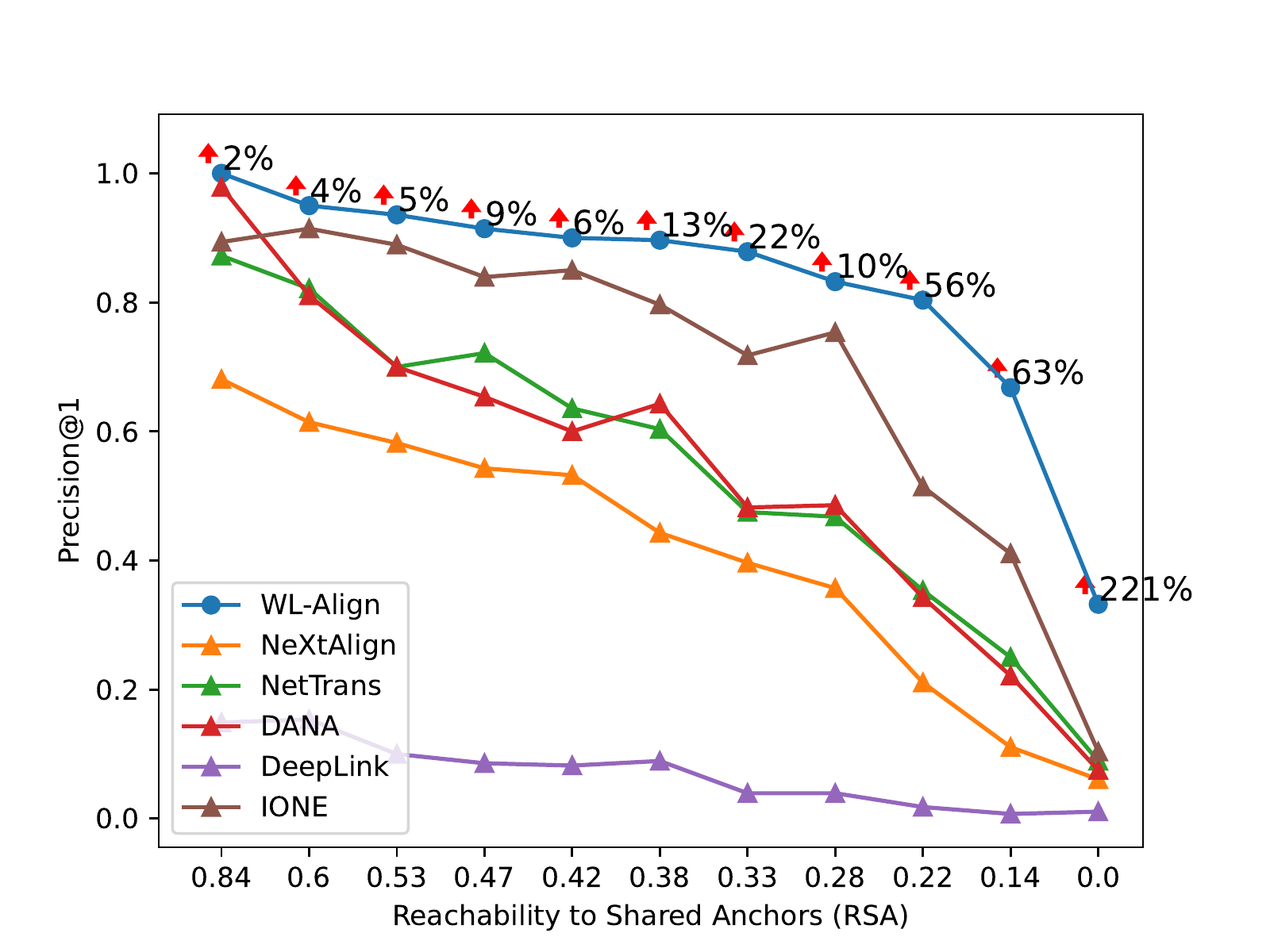}
	}
	\subfigure[Phone-Email Dataset]{
	\label{motivationpe}
		\includegraphics[width=0.3\textwidth]{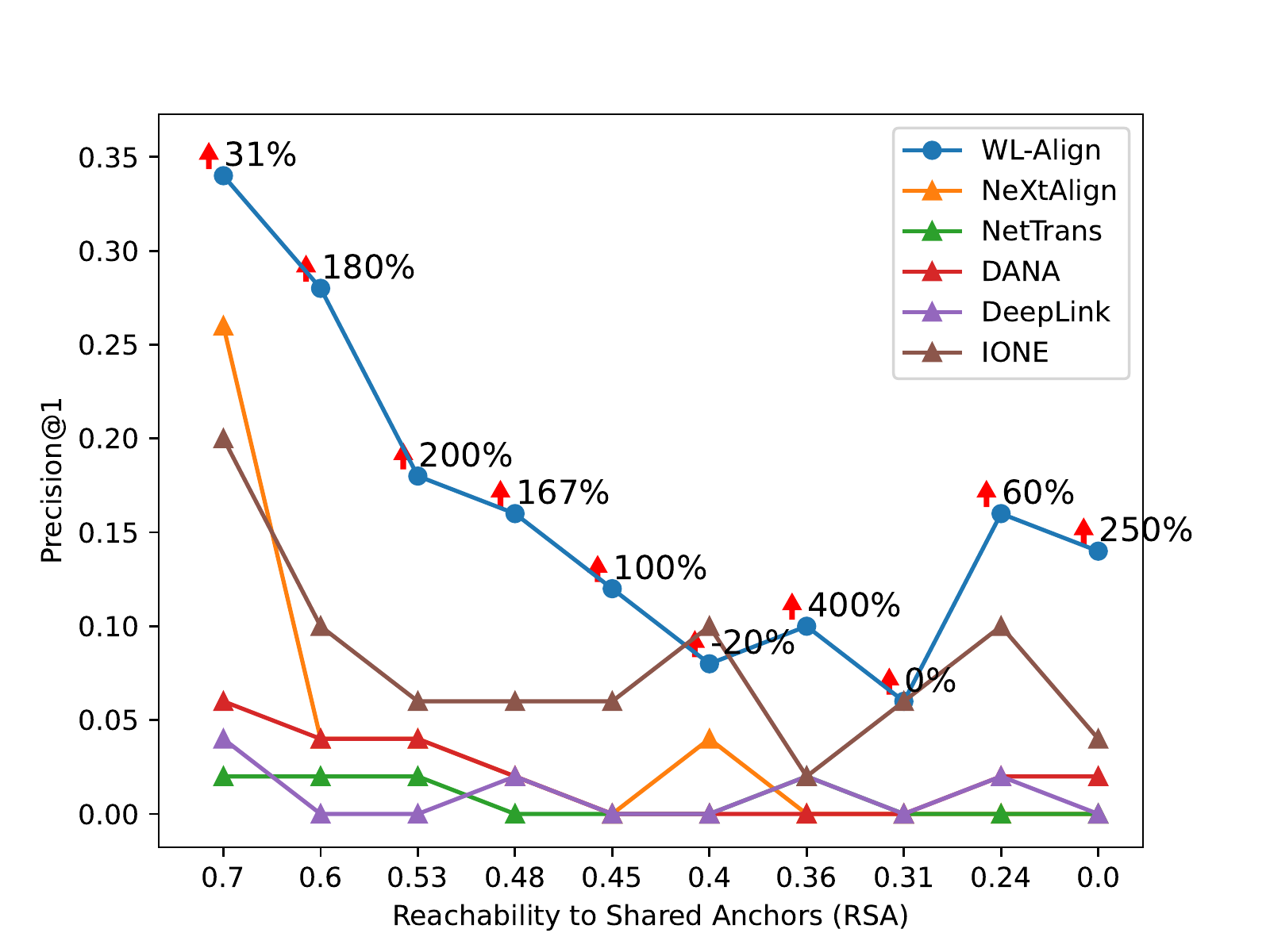}
	}
	\caption{Improvement of Precision@1 with regard to the reachability to shared anchors. The red arrows along with numbers indicate the percentage of improvement compared to the best baseline}
	\label{DSAAnalysis}
\end{figure*}

\subsection{Performance on Anchor-based Label Propagation}\label{ExpWL}

We first evaluate the robustness of the anchor-based compressed labels computed based on the proposed WL-Align algorithm without representation, which will simply be referred to as WL-Align in the section for clarity. We make use of the synthetic E-R graph and mark 20\% of its nodes as anchors. Then, we perturb the graph by randomly connecting [50\%-300\%] new edges as well as attaching [50\%-300\%] new nodes to generate pairs of graphs. 

In our experiments, we generate 18 pairs of augmented graphs and apply WL-Align to compute the anchor-based compressed labels. In particular, we evaluated WL-Align (hard) and WL-Align (soft), where the former follows the ``identical tuple'' schema to conduct the relabeling, while the latter computes the ``tuple similarity'' for the relabeling as suggested in Section \ref{sect:WL-Align}. 
As illustrated in Fig. \ref{ToySimWL}, we notice that implanting just one additional node will result in completely different labels across networks, making evaluation impractical. To alleviate the impact of this perturbation, instead of labeling within the single network, we only perform the relabeling process in WL-Align (hard) when there are identical tuples across networks.

While $Precison@N$ will be used for evaluating alignment accuracy, it however is not suitable for evaluating how well the anchor-based labels of two networks match with each other as nodes within each network can have identical compressed labels.
Instead, we use a cosine similarity and a coverage ratio. For cosine similarity, we refer to the non-augmented nodes in a pair of generated E-R graphs and represent them in the two graphs as two vectors.
Similar to the WL graph kernels~\cite{DBLP:journals/jmlr/ShervashidzeSLMB11}, each element of the vector corresponds to a specific compressed label and its value to the number of nodes with the same label. We then compute the cosine similarity of these vectors to determine how well the anchor-based labels of the two networks match each other. A larger value indicates better performance, with the highest value equal to 1 when the compressed labels of the non-augmented nodes in two graphs form an exact match across the two graphs. 
Note that some nodes may not be labeled by both WL-Align (hard) and WL-Align (soft).
Therefore, we also adopt the coverage ratio, defined as $\frac{\# labeled\ nodes}{\# total\ nodes}$, to evaluate how many non-augmented nodes can be labeled. A high coverage ratio indicates better performance. 

Fig. \ref{SynExp} illustrates the performance of WL-Align (hard) and WL-Align (soft). We observe that WL-Align (hard) drops quickly in terms of both the cosine similarity and the coverage ratio as the degree of perturbation increases. 
In contrast, WL-Align (soft) can perform much better in terms of both evaluation metrics. 
This demonstrates the robustness of WL-Align (soft) in exploring the isomorphic subgraphs of paired graphs. In the sequel, we will evaluate only WL-Align (soft) due to its superior performance and will simply refer to it as WL-Align unless otherwise specified.

\subsection{Performance on Alignment using Real-world Datasets}\label{PerOnRD}
We compare the proposed WL-Align with the baseline methods on three real-world datasets on network alignment. 
Fig. \ref{TrainingRatio50} illustrates the performance of $Precison@N$ under the 50\% training ratio. It is obvious that our proposed WL-Align outperforms all the baselines for different settings of $N$. And the improvement is particularly significant when $N$ is small. For instance, WL-Align achieves around 62.4\%, 21.8\%, and 200.0\% improvement in terms of $Precision@1$ over the most competitive models on the three datasets, respectively. This supports our claim that the anchor-based label propagation can regularize the network representation learning and benefit aligning users, especially in the exact matching scenario.

Fig. \ref{DifferentTrainingRatio} presents the performance under different training ratio settings. It is evident that: 1) For all the datasets, our proposed WL-Align outperforms all the baselines under different settings of training ratio; 2) With the increase of the training ratio, the performance of all the models improves accordingly, indicating that the known anchors provide key information to boost the performance of the alignment models; 3) We observe significant improvement achieved by WL-Align on both the Twitter-Foursquare and Phone-Email datasets, but with only limited improvement on the ACM-DBLP dataset. We further check that the number of connected components in the six networks in the datasets are 1, 2, 234, 9, 2, and 1 for Twitter, Foursquare, ACM, DBLP, Phone, and Email networks, respectively. 
We believe the limited improvement achieved on the ACM-DBLP dataset is due to the relatively large number of connected components in the ACM network which hinders the propagation of the hash labels and thus limits the contribution of Eq. \eqref{loss_label}. But still, when given a rather small number of known anchors (a small training ratio), the superiority of WL-Align is more evident, showing its effectiveness in exploiting the information carried by the anchors.  

To further investigate how the topological range between nodes affects the alignment performance, we define a metric called \textit{reachability to shared anchors (RSA)} to evaluate the alignment performance of $Precision@1$ with respect to each potential anchor pair ($v_i^s$, $v_j^t$), shown in Eq. \eqref{RSA}. 
\begin{equation}
    RSA=\sum_{i=1}^3\lambda^{(i-1)}\frac{2\times|Anchor_{i-hop}|}{|neighbour_{i-hop}(v_i^s)|+|neighbour_{i-hop}(v_j^t)|}
    \label{RSA}
\end{equation}
where $|Anchor_{i-hop}|$ is the number of shared anchors among the $i-th$ hop neighbours of $v_i^s$ and $v_j^t$.  $|neighbour_{i-hop}(v_i^s)|$ and $|neighbour_{i-hop}(v_j^t)|$ are the numbers of the $i-th$ hop neighbours of $v_i^s$ and $v_j^t$, respectively. $\lambda$ is a discount factor and is set to 0.5. A larger RSA indicates potential anchors can reach anchors more easily. According to the calculated RSAs of all potential anchor pairs for each dataset, we divide them into 10 parts equally. Fig. \ref{DSAAnalysis} shows $Precision@1$ performance with respect to RSA for each dataset. It is evident that when RSA is large, i.e., the potential anchors share many anchors within small hops, all models present their competitively good performance. In contrast, when RSA decreases, we observe the significant outperformance achieved by WL-Align. It indicates that WL-Align can better utilize the discriminant information of long-range anchors, benefiting representation learning for aligning users across networks.

\begin{figure}
	\centering
	\includegraphics[width=0.45\textwidth]{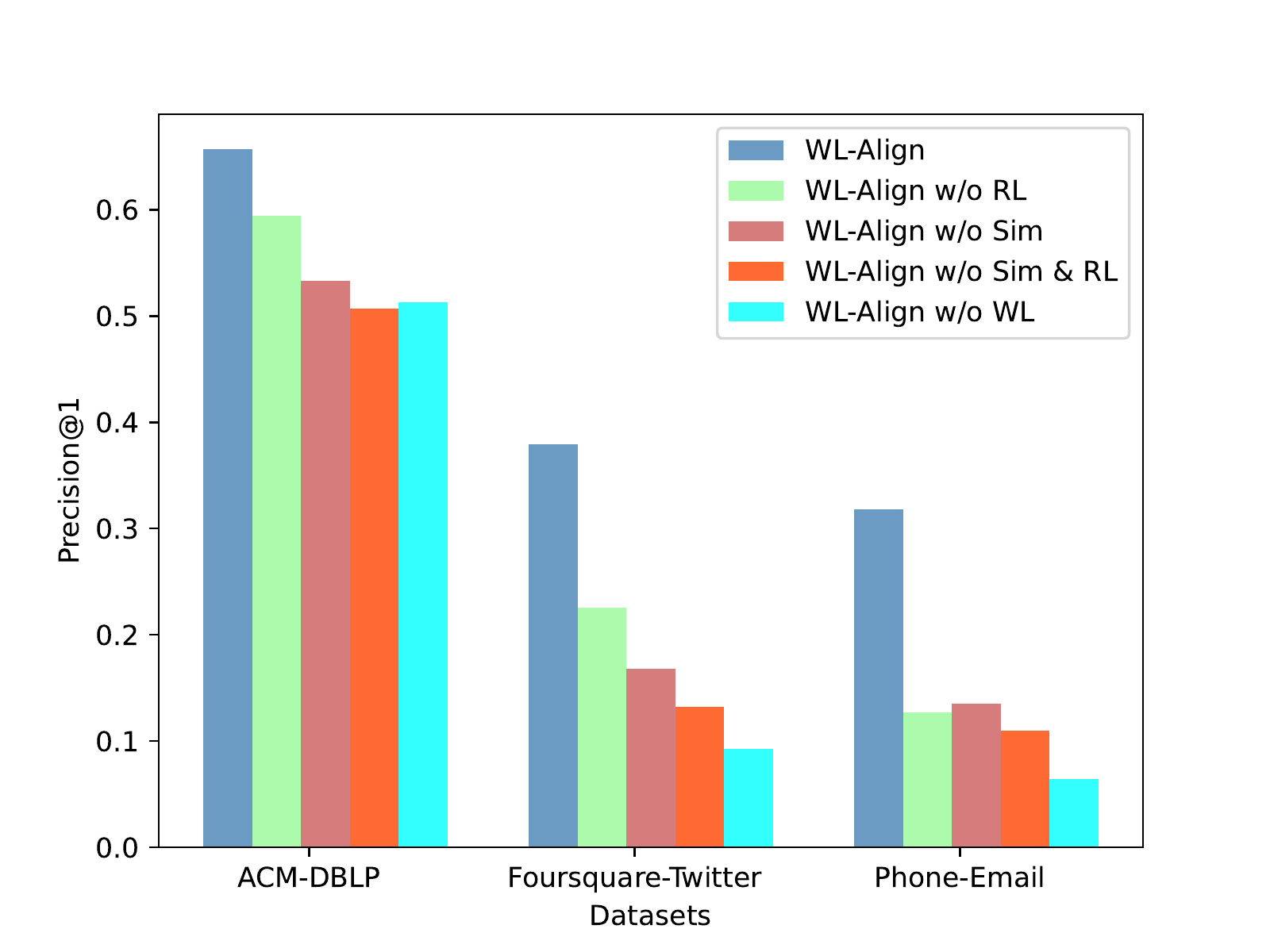}
	\caption{Ablation study with Training Ratio 50\%}
	\label{ablation50}
\end{figure}

\subsection{Ablation Study}
We validate the effectiveness of different components in the proposed WL-Align model through an experimental comparison between the variants of \textit{WL-Align}, namely, \textit{WL-Align w/o RL}, \textit{WL-Align w/o Sim}, \textit{WL-Align w/o Sim \& RL} and \textit{WL-Align w/o WL}.
\begin{enumerate}
    \item \textit{WL-Align w/o RL} contains only the ``anchor-based label propagation'' module and directly utilizes the similarity between tuples which is computed according to Eq. \eqref{wl_similarity} to perform the alignment. It can be considered as the supervised information for optimizing the loss $\mathcal{L}_{label}$.
    \item \textit{WL-Align w/o WL} only contains the ``representation learning'' module, where $\mathcal{L}_{context}$ is leveraged to obtain the node representations.  
    \item \textit{WL-Align w/o Sim} differs from \textit{WL-Align} by utilizing an injective hashing based on identical tuples in the conventional WL Test to relabel nodes. 
    \item \textit{WL-Align w/o Sim \& RL} performs the alignment by simply using the hash labels induced by \textit{WL-Align w/o Sim} without the representation learning module. 
\end{enumerate}

Fig. \ref{ablation50} illustrates model performance in terms of $Precison@1$ on the three datasets. We can observe that: 1) Models incorporated with the ``anchor-based label propagation'' module show an overwhelming relative advantage compared to \textit{WL-Align w/o WL}, indicating the importance of our proposed idea of exploiting the anchor-based compressed labels; 2) The use of similarity-based hashing is able to further boost the model performance as seen from the significant performance gaps between \textit{WL-Align} and \textit{WL-Align w/o Sim}, \textit{WL-Align w/o RL}, and \textit{WL-Align w/o Sim\& RL}; 3) Compared to all the variants, \textit{WL-Align} performs best due to both the similarity-based label hashing and structure-preserving representation learning.

\subsection{Visualization of Representation Space}

\begin{figure*}[!t] \centering
    \subfigure[WL-Align]{\includegraphics[width=0.31\textwidth]{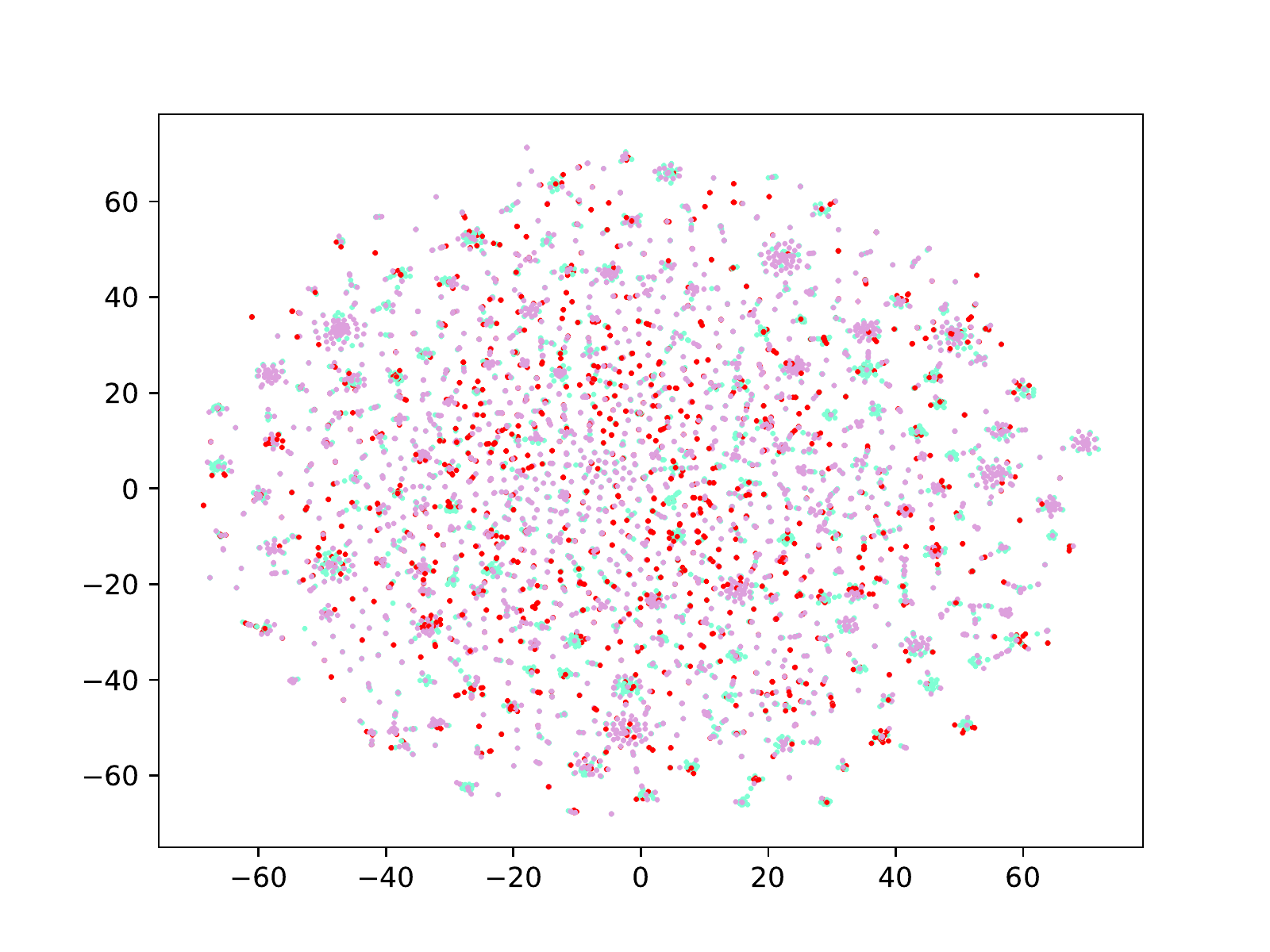}
	}
	\subfigure[IONE] { \includegraphics[width=0.31\textwidth]{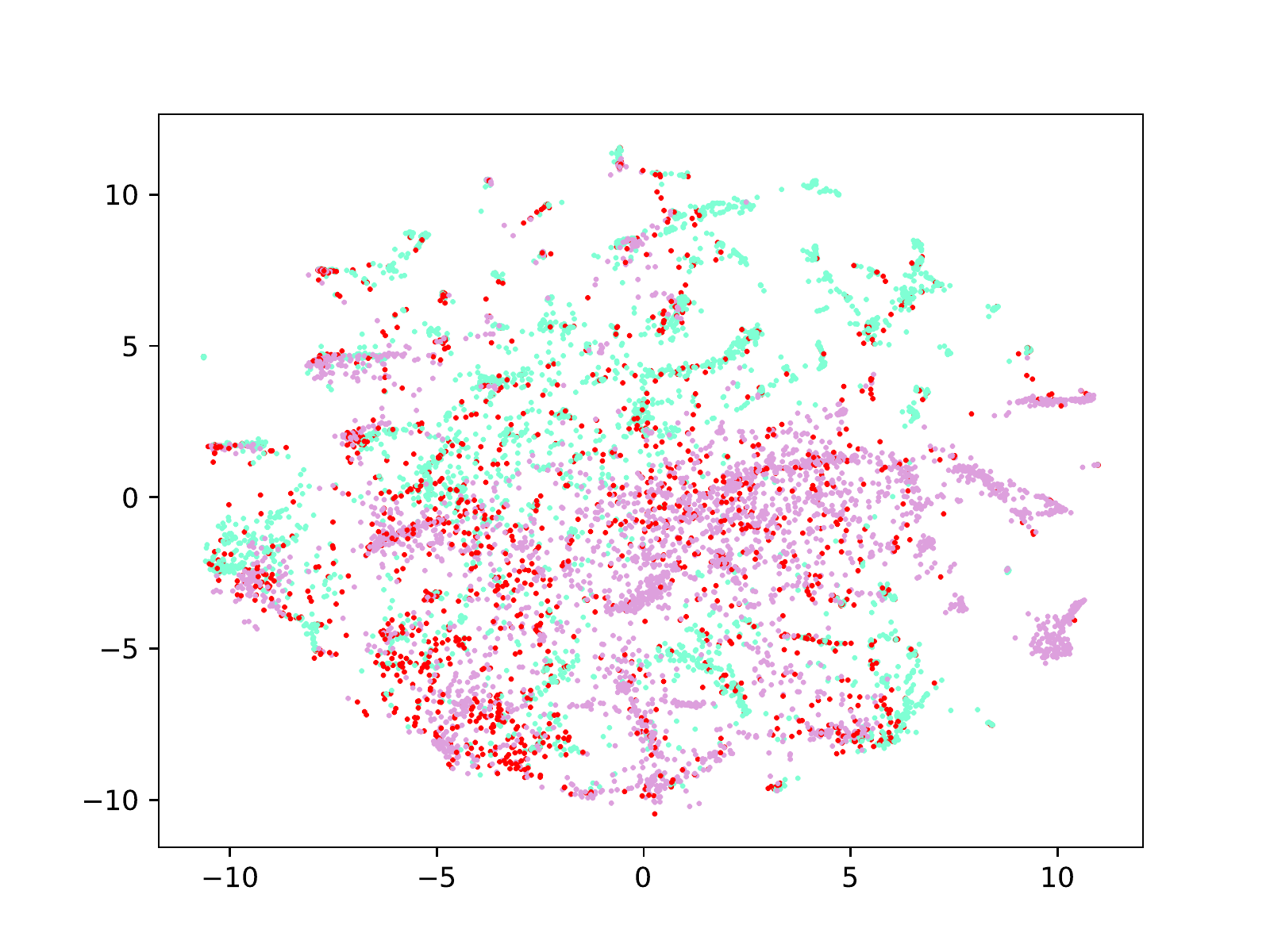}
	}
	\subfigure[DEEPLINK]{\includegraphics[width=0.31\textwidth]{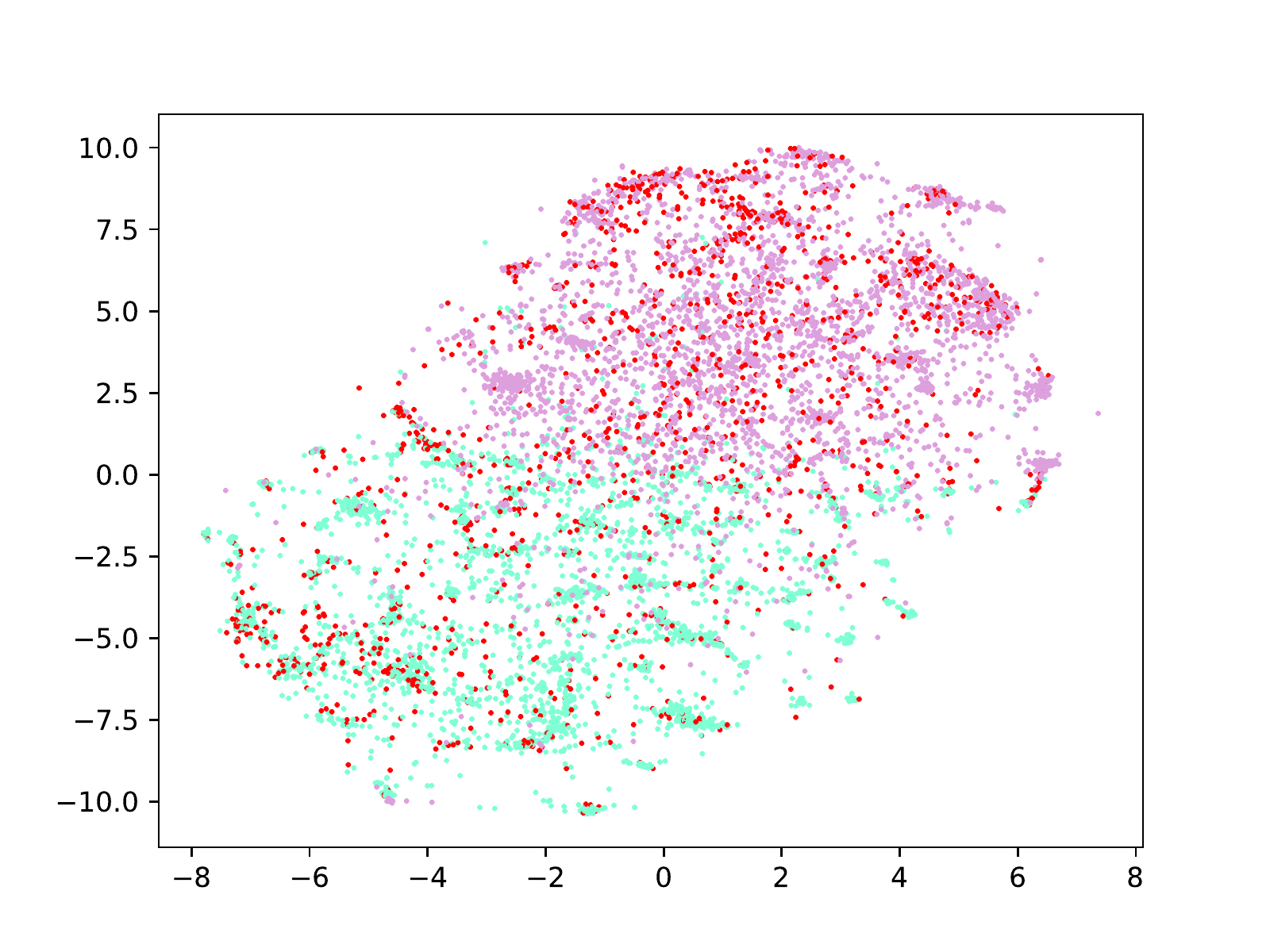}
	}
    \subfigure[DANA]{\includegraphics[width=0.31\textwidth]{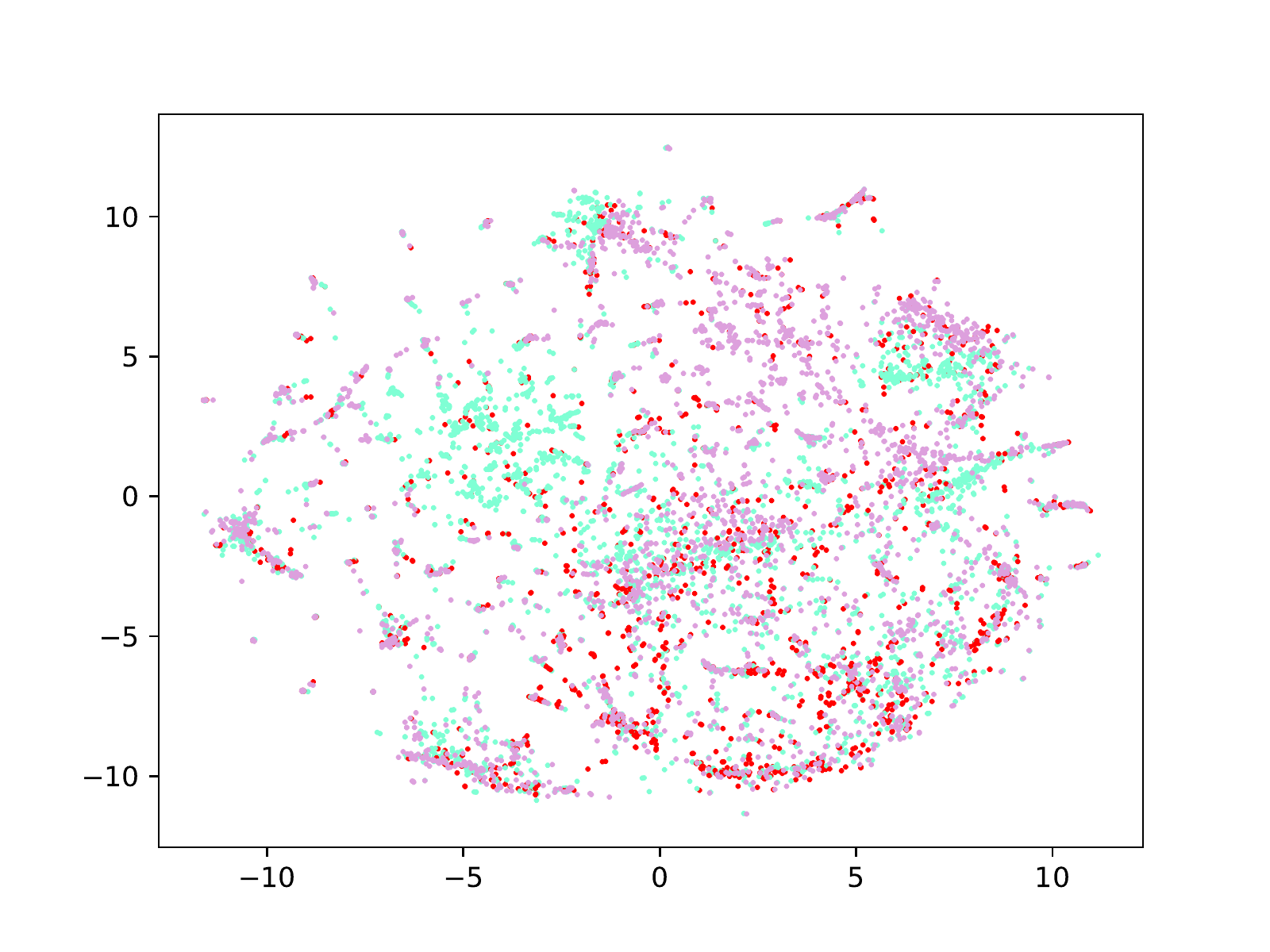}
    }
	\subfigure[NetTrans] {\includegraphics[width=0.31\textwidth]{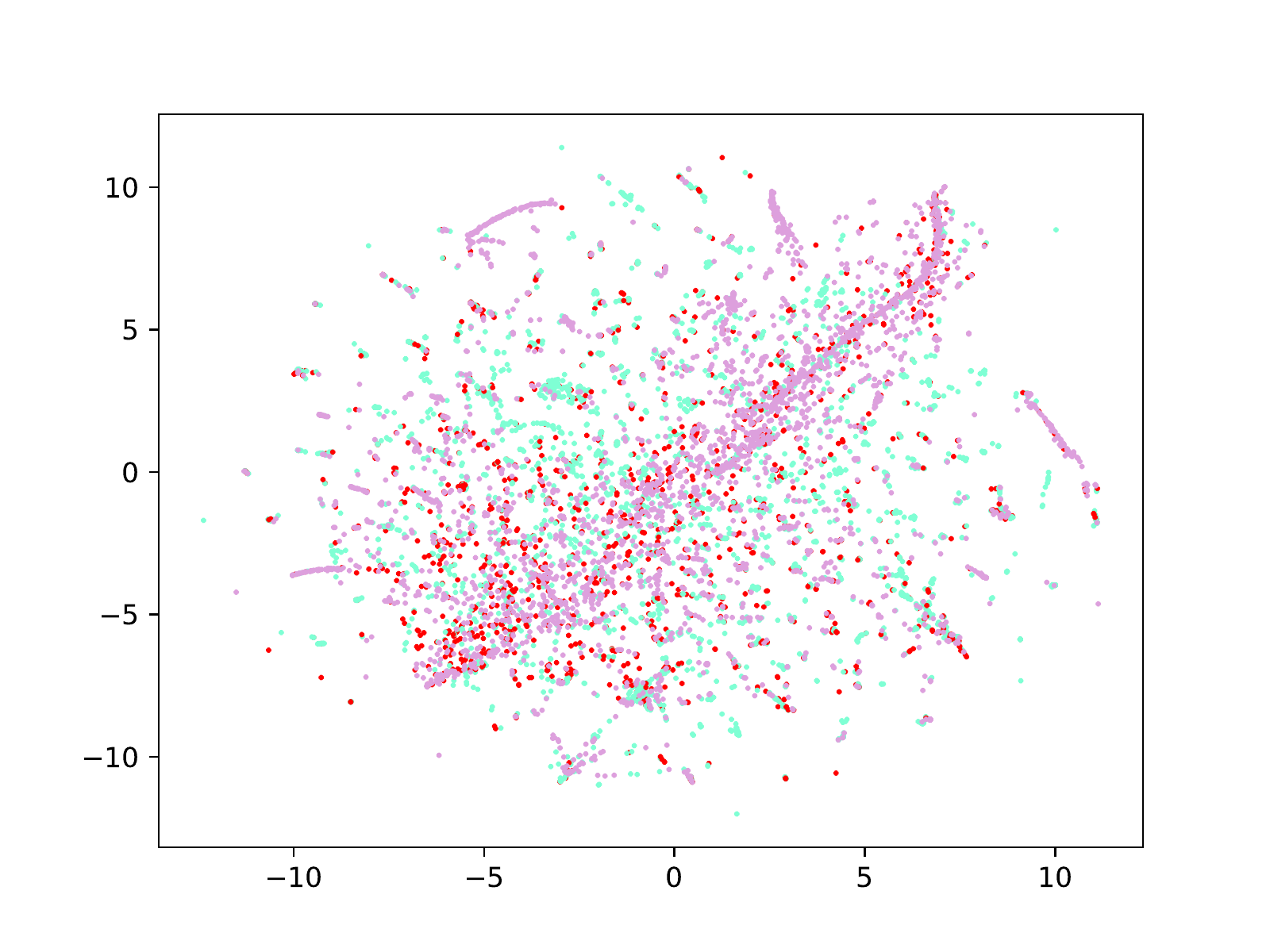}
	}
	\subfigure[NeXtAlign]{\includegraphics[width=0.31\textwidth]{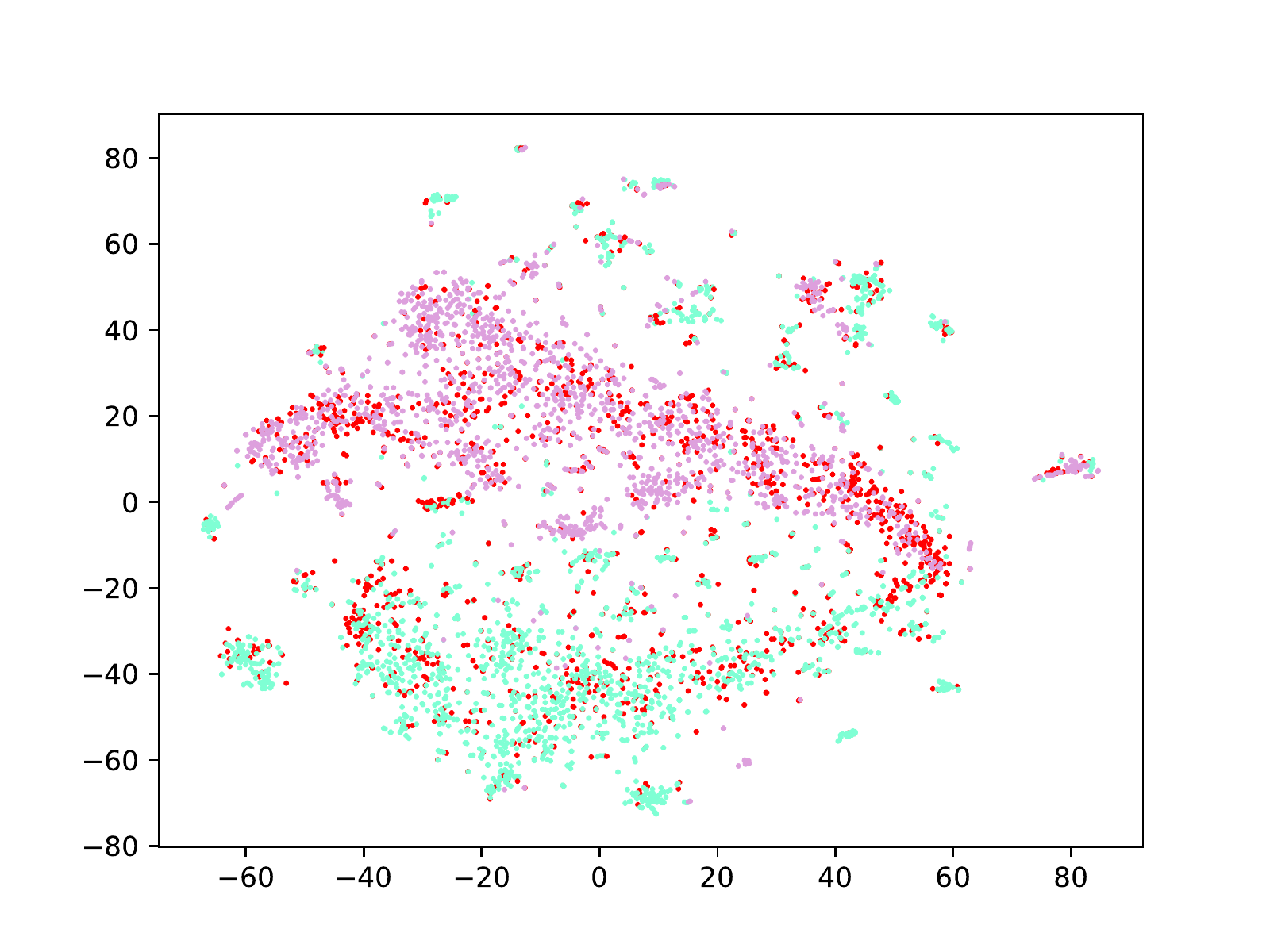}
	}
	\caption{Visualization of Representation Space (Twitter Foursquare)}
	\label{OBS-FT}
\end{figure*}

Fig. \ref{OBS-FT} visualizes the learned representations over the Twitter-Foursquare dataset for different methods using t-SNE \cite{van2008visualizing}. The red dots represent the potential anchors to be aligned, whereas the pink and green dots represent nodes in the respective network.
Compared to the dense structures learned by other baselines, the node embeddings obtained by WL-Align are spatially organized into a large number of small groups evenly distributed in the embedding space. This is consistent with what we expect where nodes with the same compressed labels across the social networks should be close and those with different labels should be far apart in the embedding space. This accounts for the better performance achieved by WL-Align.

\begin{figure}[H]
	\centering
	\includegraphics[width=0.5\textwidth]{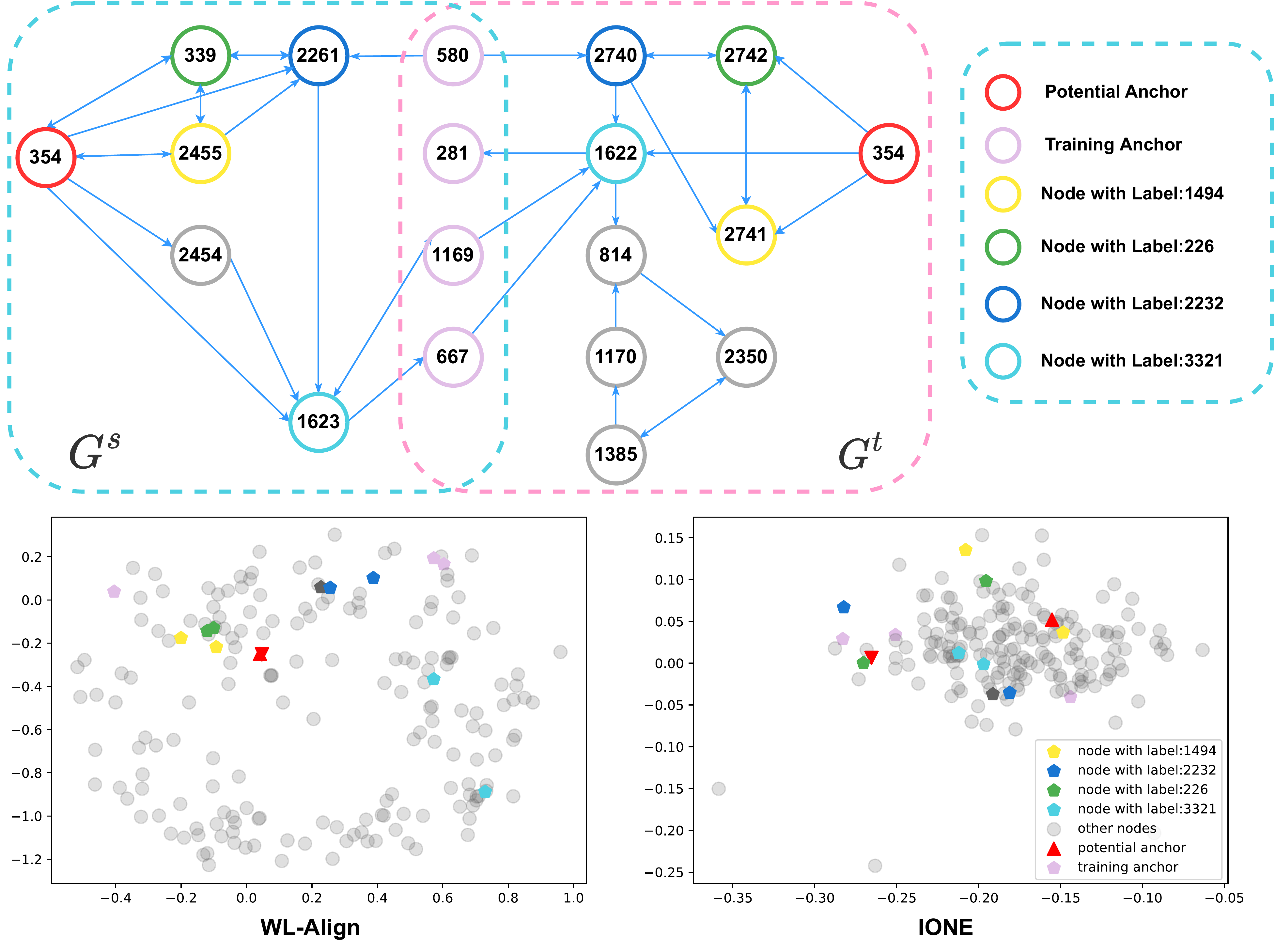}
	\caption{Micro-Level Case Study}
	\label{CaseStudy}
\end{figure}

We further take a more microscopic view and provide a case study where we select a sub-network formed by a potential anchor pair (nodes 354) in the Twitter-Foursquare dataset, where nodes in different networks are connected by the anchors in pink and the number inside each node represents its index. Note that the node index is not used as a feature in our WL-Align to avoid data leakage. Fig. \ref{CaseStudy} illustrates the t-SNE based visualizations of embeddings learned by WL-Align and IONE, respectively. We notice that the IONE model, which learns embeddings via preserving second-order proximity, obtains a rather dense embedding space, resulting in indistinguishable representations for nodes 354 across networks. Benefiting from the propagation and the label compressing, we observe that nodes 354 in the separated networks share nodes with the same labels, which are Label:226, Label:2232, and Label:3321. The similar context nodes making nodes 354 are close to each other in the representation space. All nodes are well organized in the learned space according to their compressed labels. This is consistent with our initial motivation.

\subsection{Training Configuration Analysis}

\begin{figure*}[!t] \centering
    \subfigure[Twitter-Foursquare Dataset]{
	\label{TFIter}
		\includegraphics[width=0.3\textwidth]{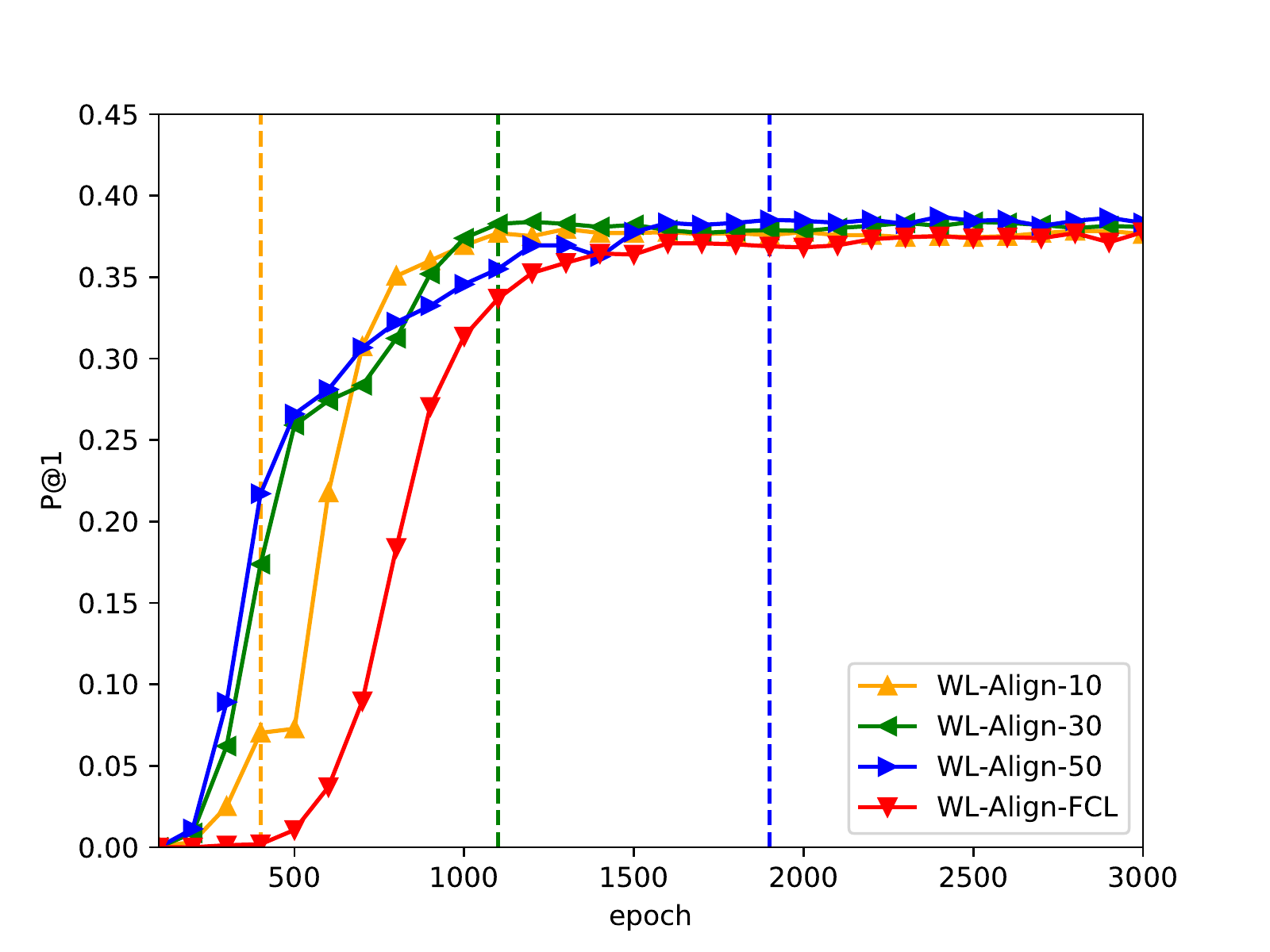}
	}
	\subfigure[ACM-DBLP Dataset] { \label{ADIter}
		\includegraphics[width=0.3\textwidth]{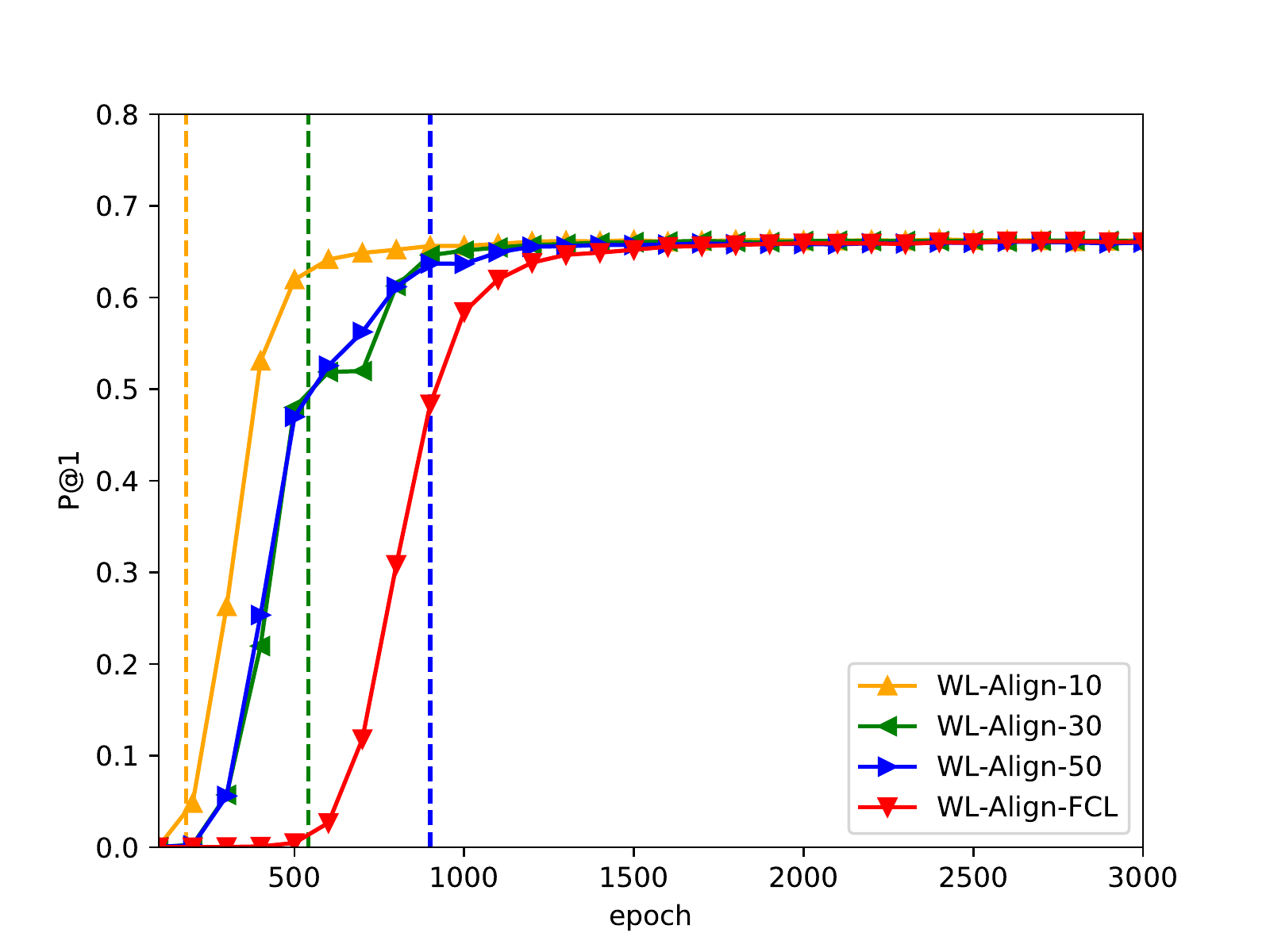}
	}
	\subfigure[Phone-Email Dataset]{
	\label{PEIter}
		\includegraphics[width=0.3\textwidth]{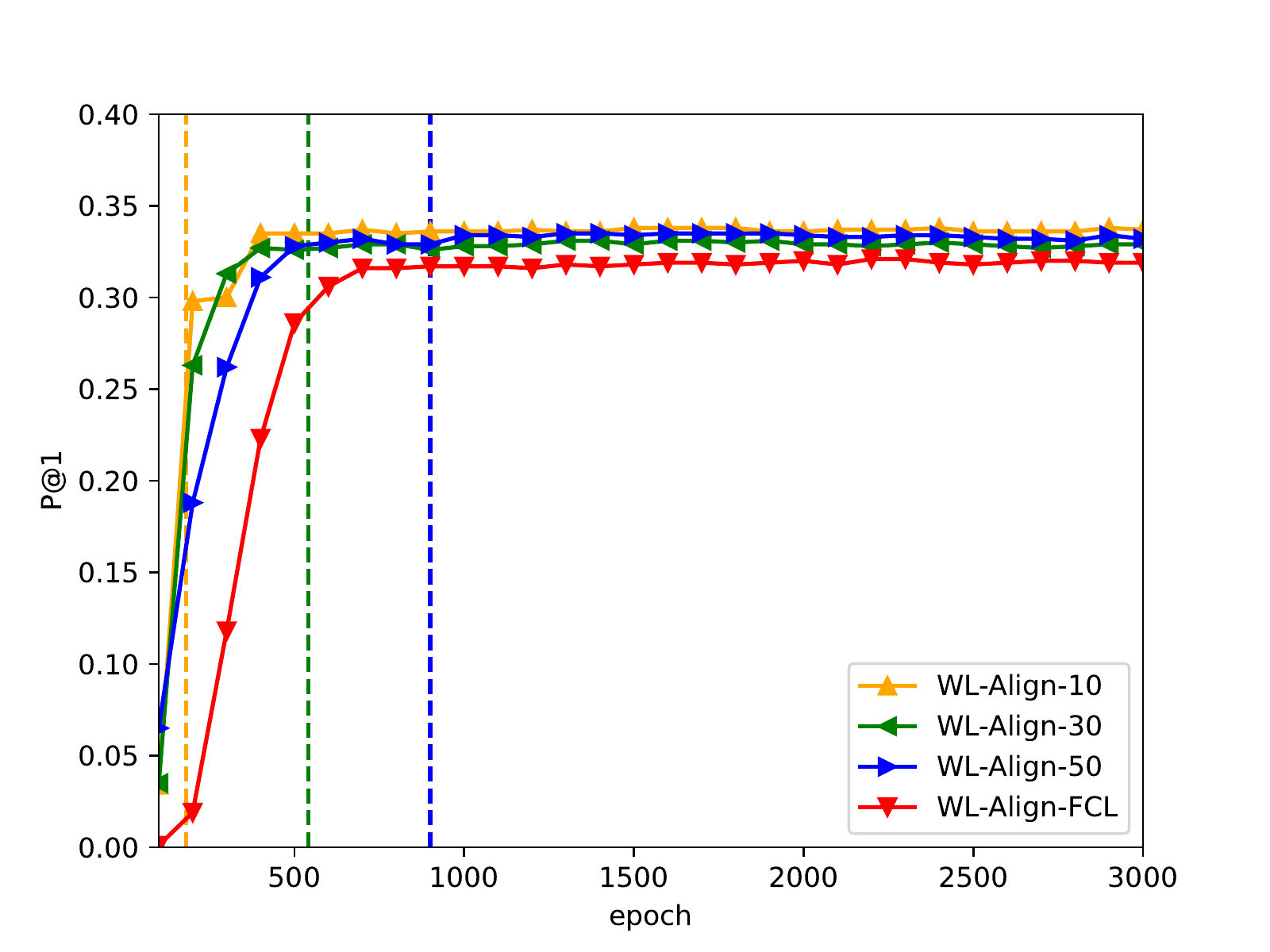}
	}
	\caption{Performance of WL-Align at Different Training Settings.}
	\label{Iteration}
\end{figure*}

The training of WL-Align proceeds in two alternating phases: 1) one iteration of label propagation, and 2) $E$ iterations of representation learning (see Algorithm \ref{algo}). The two phases are interleaved until they converge. We conduct experiments with different settings of $E$ for training WL-Align to observe the sensitivity of the setting to the overall performance. Fig. \ref{Iteration} shows the experimental results on the three datasets. The $x$ axis denotes the number of iterations of the representation learning conducted. The colored solid curves represent the results obtained by WL-Align-10/30/50 and WL-Align-FCL respectively. WL-Align-10/30/50 correspond to the settings with the hyperparameter $E$ set as 10/30/50. The vertical dashed lines with different colors indicate the number of iterations elapsed for different settings of $E$ when the label propagation process reaches its convergence. WL-Align-FCL refers to the setting with the label propagation carried out first until the compressed label set does not change, and then followed by 3,000 iterations of representation learning.   

We have the following observations: 1) WL-Align-10/30/50 show higher convergence rates as compared to WL-Align-FCL. This shows the effectiveness of alternating the two phrases of training as the ``imperfect'' labels are informative to some extent for guiding the training. 2) When $E$ is set to 10, the proposed model shows better convergence performance for the ACM-DBLP and Phone-Email datasets, while the best performance is achieved on the Twitter-Foursquare dataset for a larger setting of $E$. Our conjecture is that for a dataset with relatively small networks, a small number of iterations is sufficient for the gradient descent to find the proper representations under the guidance of the current compressed labels. Larger networks inherently require more iterations to achieve good representations. Note that the networks in the ACM-DBLP dataset consist of a large number of connected components. Thus, WL-Align is essentially working on a group of small-sized networks, despite the number of nodes in ACM-DBLP being the largest among all three datasets. 3) There exists sharp performance improvement for WL-Align-10 on the Twitter-Foursquare dataset after the anchor-based compressed labels reached their convergence though with a small setting of $E$, indicating that the anchor-based labels provide more informative signals to guide the representation learning.

\section{Conclusion}
In this paper, to address the ``over-smoothing'' issue when the GRL-based alignment models encounter long-range anchors, we propose a novel methodology for the network alignment task called WL-Align, where an across-network Weisfeiler-Lehman relabeling process is developed to regularize the node representation learning. By performing anchor-based propagation and similarity-based hashing, nodes can be effectively characterized by the compressed hash labels which implicitly show how the anchors are connected to them. The node representations can then be learned via preserving the label and the structure proximity simultaneously. The experimental results obtained from the real-world and synthetic datasets demonstrate that our proposed model can effectively learn a well-organized representation space, and can outperform the state-of-the-art user alignment methods, especially for the exact-matching scenario. For future work, we are interested in seeing if the consideration of the WL isomorphism test in WL-Align can provide us further insights on analyzing the performance upper bound of the structure-based alignment model.

\ifCLASSOPTIONcompsoc
  \section*{Acknowledgments}
\else
  \section*{Acknowledgment}
\fi

The work is partially supported by National Natural Science Foundation of China (61936001, 61806031), and in part by the Natural Science Foundation of Chongqing (cstc2019jcyj-cxttX0002, cstc2020jcyj-msxmX0943), and in part by the key cooperation project of Chongqing  Municipal Education Commission (HZ2021008), and in part by Science and Technology Research Program of Chongqing Municipal Education Commission (KJQN202100629, KJQN202001901), and in part by Doctoral Innovation Talent Program of Chongqing University of Posts and Telecommunications (BYJS202118). This work is partially done when Li Liu works at Hong Kong Baptist University supported by the Hong Kong Scholars program (XJ2020054). 

\ifCLASSOPTIONcaptionsoff
  \newpage
\fi



\bibliographystyle{IEEEtran}
\bibliography{IEEEabrv,WLAlign}%

%
\vspace{-1.5CM}
\begin{IEEEbiography}[{\includegraphics[width=1in,height=1.25in,clip,keepaspectratio]{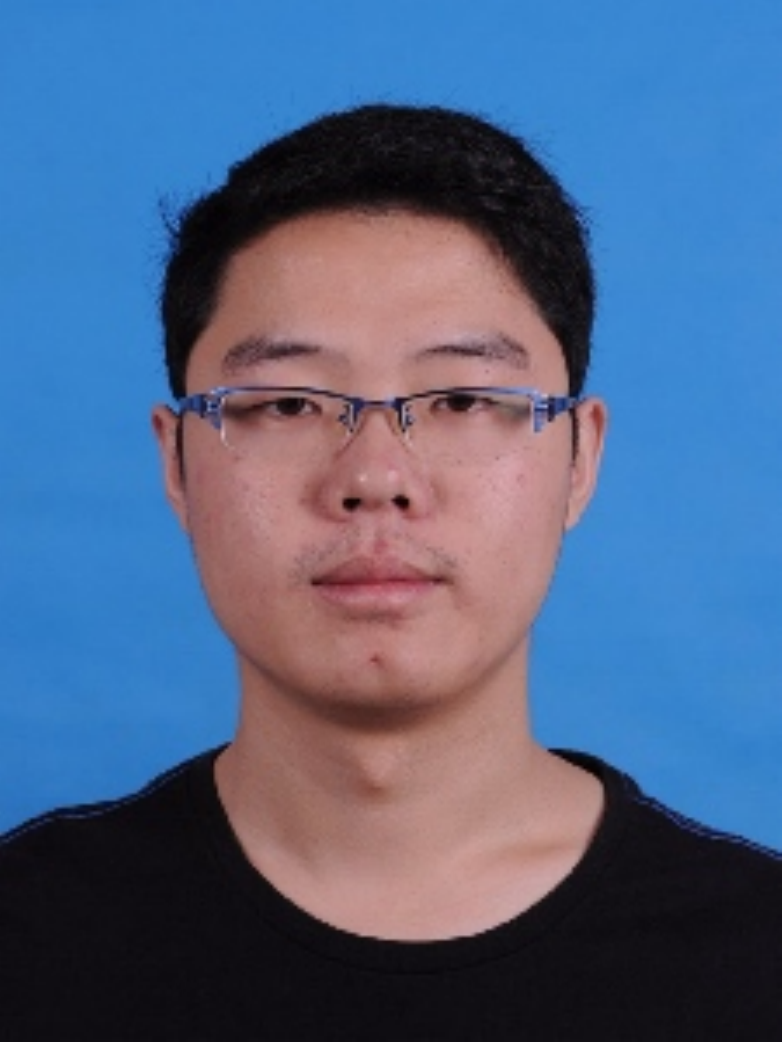}}]{Li Liu}
received the  Ph.D. in computer science from Beijing Institute of Technology in 2016, the M.E. in computer science from Kunming University of Science and Technology in 2012 and the B.B.A. in Information management and information system from Chongqing University of Posts and Telecommunications in 2009. He was a visiting student at Hong Kong Baptist University during 2016. He is currently an Associate Professor in Chongqing University of Posts and Telecommunications. His research interests include web mining and social computing.
\end{IEEEbiography}
\vspace{-1.5CM}

\begin{IEEEbiography}[{\includegraphics[width=1in,height=1.25in,clip,keepaspectratio]{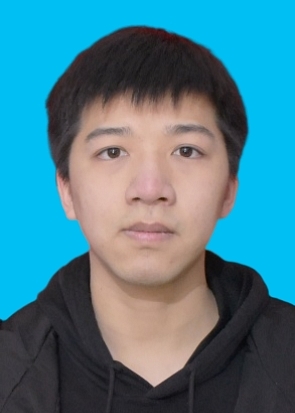}}]{Penggang Chen} received his B.S. degree in Computer science from the Chongqing University of Posts and Telecommunications, China, in 2020. He is currently pursuing a M.S. degree in computer technology at the Chongqing University of Posts and Telecommunications in Chongqing, China. His research interests include social network and data mining.
\end{IEEEbiography}
\vspace{-1.5CM}


\begin{IEEEbiography}[{\includegraphics[width=1in,height=1.25in,clip,keepaspectratio]{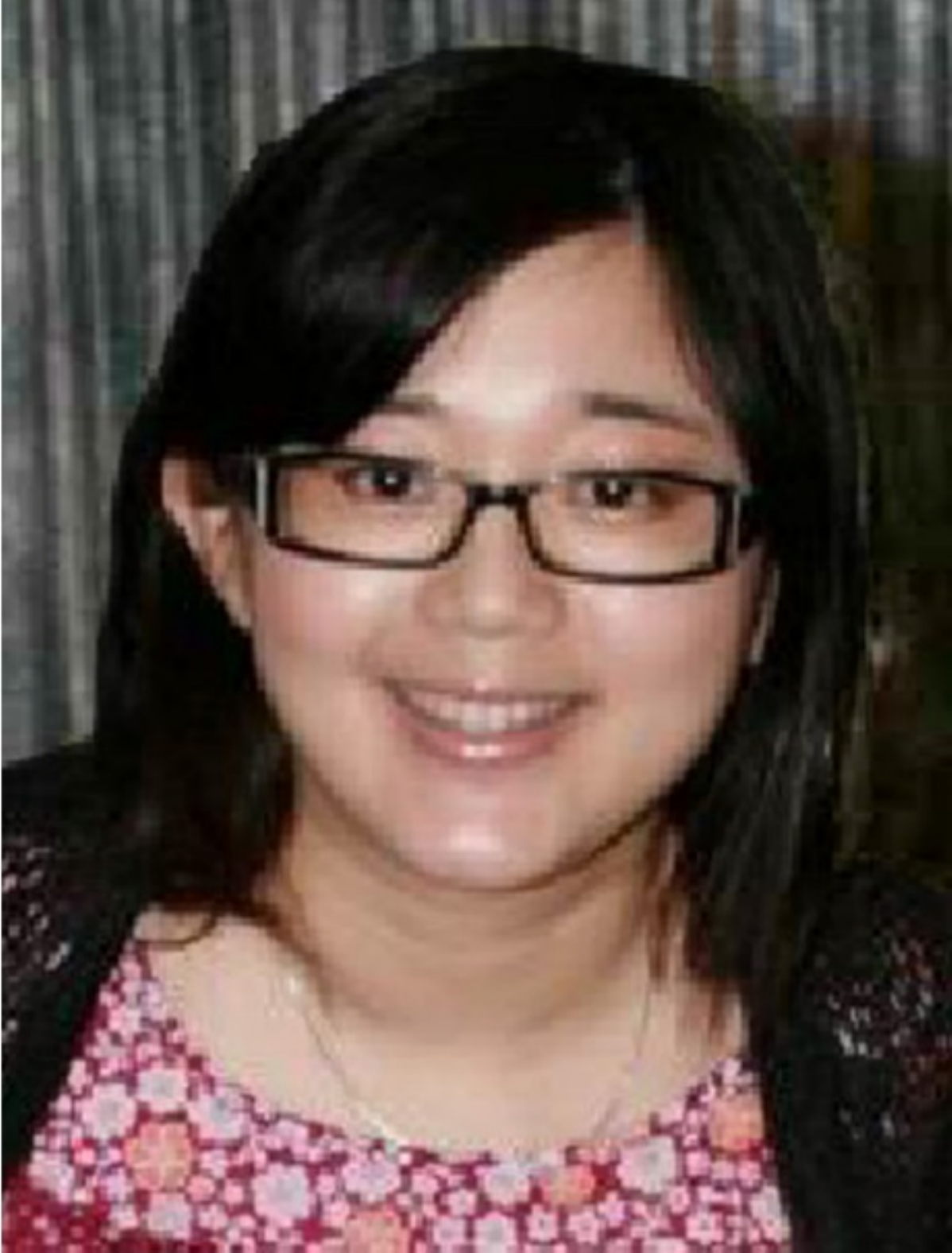}}]{Xin Li}
is currently an Associate Professor in the School of Computer Science at Beijing Institute of Technology, China.  She received the B.Sc. and M.Sc degrees in Computer Science from Jilin University  China, and the Ph.D. degree in Computer Science at Hong Kong Baptist University. Her research focuses on the development of algorithms for representation learning, reasoning under uncertainty and machine learning with application to Natural Language Processing, Recommender Systems, and Robotics.
\end{IEEEbiography}

\begin{IEEEbiography}[{\includegraphics[width=1in,height=1.25in,clip,keepaspectratio]{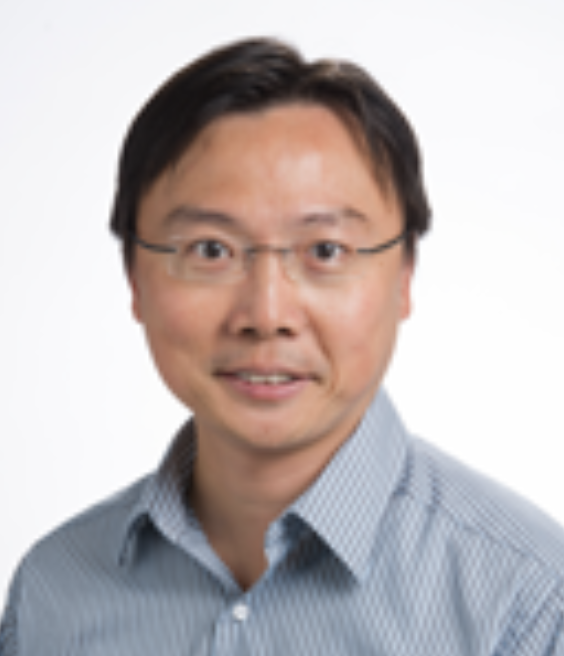}}]{William K. Cheung} received the Ph.D. degree in computer science from the Hong Kong University of Science and Technology in Hong Kong in 1999. He is currently Professor of the Department of Computer Science, Hong Kong Baptist University, Hong Kong. His current research interests include artificial intelligence, data mining, collaborative information filtering, social network analysis, and healthcare informatics. He has served as the Co-Chairs and Program Committee Members for a number of international conferences and workshops, as well as Guest Editors of journals on areas including artificial intelligence, Web intelligence, data mining, Web services, e-commerce technologies, and health informatics. From 2002-2018, he was on the Editorial Board of the IEEE Intelligent Informatics Bulletin. He is currently a Track Editor of Web Intelligence Journal and an Associate Editor of Journal of Health Information Research, and Network Modeling and Analysis for Health Informatics and Bioinformatics.
\end{IEEEbiography}

\begin{IEEEbiography}[{\includegraphics[width=1in,height=1.25in,clip,keepaspectratio]{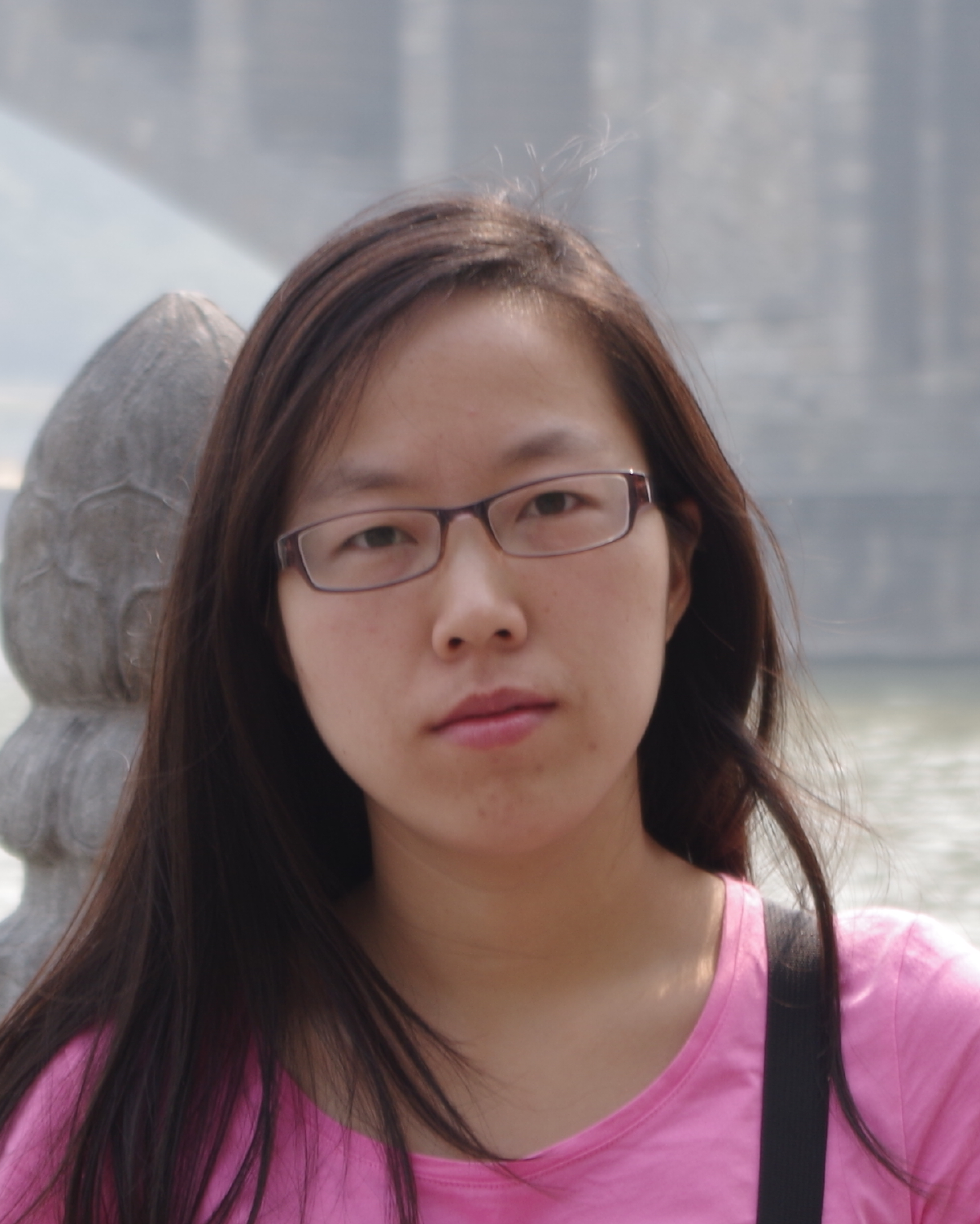}}]
{Youmin Zhang} received M.E. in control engineering from Kunming University of Science and Technology in 2013 and the B.E. in automation from University of Jinan 2009. She is currently pursuing a Ph.D. degree in computer technology at the Chongqing
University of Posts and Telecommunications in
Chongqing, China. She is currently an Assistant Professor in Chongqing Institute of Engineering. Her research interests include social computing and knowledge graph mining.
\end{IEEEbiography}

\begin{IEEEbiography}[{\includegraphics[width=1in,height=1.25in,clip,keepaspectratio]{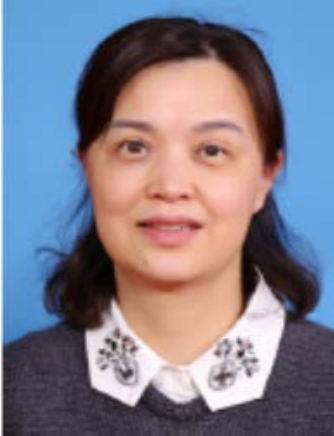}}]{Qun Liu} received her B.S. degree from Xi’An Jiaotong University in China in 1991, and the M.S. degree from Wuhan University in China in 2002, and the Ph.D from Chongqing University in China in 2008. She is currently a Professor with Chongqing University of Posts and Telecommunications. Her current research interests include complex and intelligent systems, neural networks and intelligent information processing.
\end{IEEEbiography}

\begin{IEEEbiography}[{\includegraphics[width=1in,height=1.25in,clip,keepaspectratio]{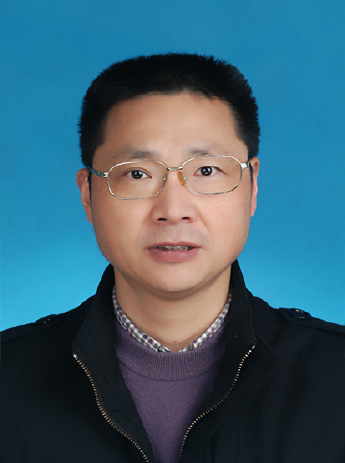}}]{Guoyin Wang}(SM’03) received the B.S., M.S., and Ph.D. degrees from Xi’an Jiaotong University, Xian, China, in 1992, 1994, and 1996, respectively. He was at the University of North Texas, and the University of Regina, Canada, as a visiting scholar during 1998-1999. Since 1996, he has been at the Chongqing University of Posts and Telecommunications, where he is currently a professor, the director of the Chongqing Key Laboratory of Computational Intelligence, the Vice-President of the University and the dean of the School of Graduate. He was appointed as the director of the Institute of Electronic Information Technology, Chongqing Institute of Green and Intelligent Technology, CAS, China, in 2011. He is the author of over 10 books, the editor of dozens of proceedings of international and national conferences, and has more than 300 reviewed research publications. His research interests include rough sets, granular computing, knowledge technology, data mining, neural network, and cognitive computing, etc. Dr. Wang was the President of International Rough Set Society (IRSS) 2014-2017. He is a Vice-President of the Chinese Association for Artificial Intelligence (CAAI), and a council member of the China Computer Federation (CCF).

\end{IEEEbiography}



\clearpage
\onecolumn
\appendices
\section{Detailed Process of WL-Align (hard) in Perturbed Networks}\label{appendix:WLHard}
\begin{figure*}
	\centering
	\includegraphics[width=0.95\textwidth]{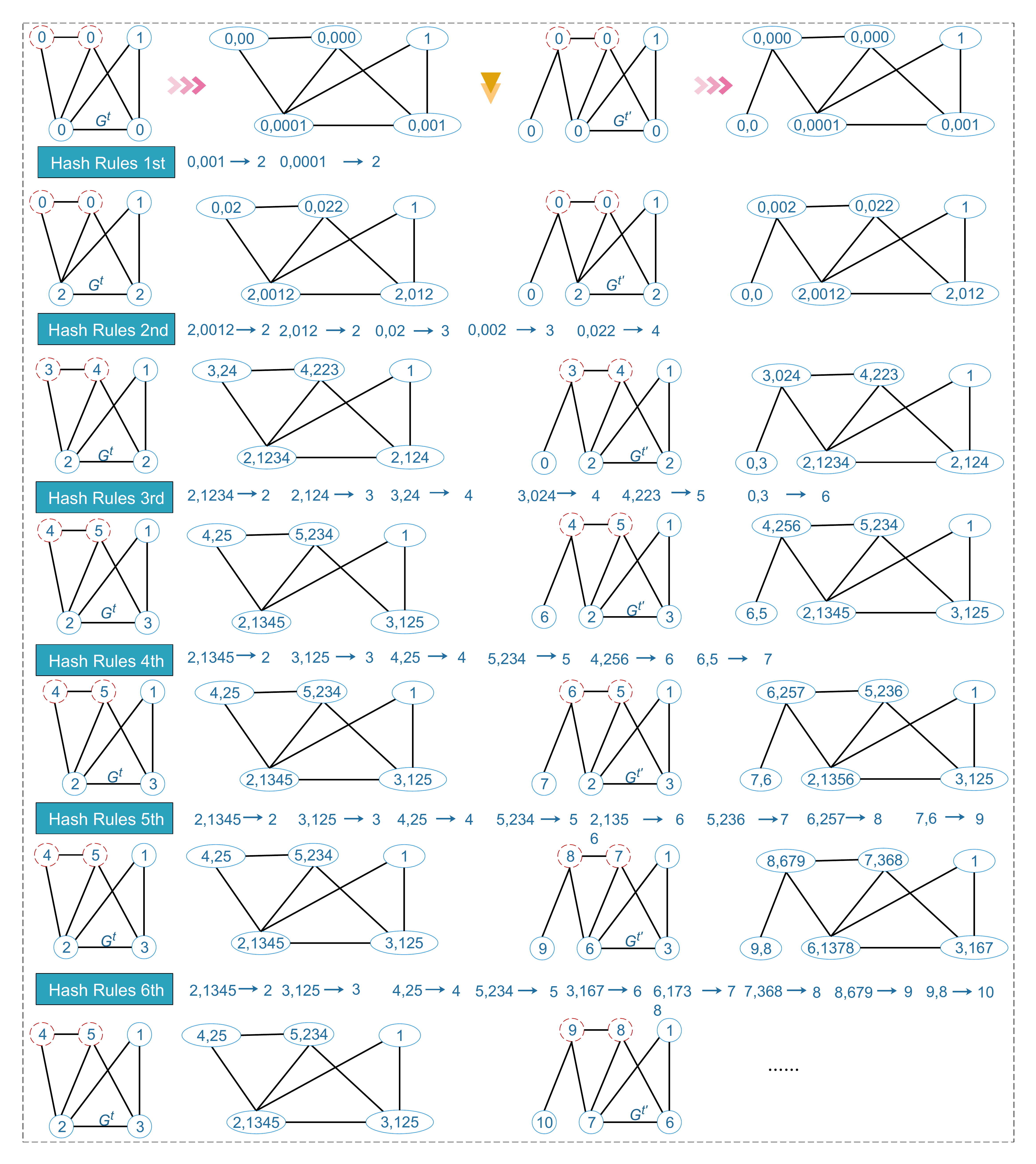}
	\caption{The Relabeling and Propagation Process in the Perturbed Network Using WL-Align (hard). Note the "0" label is ignored in the hash rules as it does not provide supervised information for learning unified representations.}
	\label{WLHARD}
\end{figure*}

The detailed process of WL-Align (hard) is illustrated in Fig. \ref{WLHARD}. The common rules are identical to ensure the propagation and label compression (using injective hashing function) across the networks can be conducted in the unified space.

\end{document}